\documentclass[aip,jcp,reprint]{revtex4-1}

\usepackage{graphicx}
\usepackage{amsmath, amsthm}
\usepackage[dvipsnames]{xcolor}
\usepackage{subfigure}
\usepackage{enumerate}
\usepackage{lipsum}
\usepackage[normalem]{ulem}

\begin{document}

\title{Multivalent ``Attacker \& Guard'' Strategy for Targeting Surfaces with Low Receptor Density}
\author{Nicholas B. Tito}
	\affiliation{Department of Applied Physics, Eindhoven University of Technology, PO Box 513, 5600 MB, Eindhoven, The Netherlands \\ Institute for Complex Molecular Systems, Eindhoven University of Technology, PO Box 513, 5600 MB, Eindhoven, The Netherlands}
	\email{nicholas.b.tito@gmail.com}

\date{\today}
\begin{abstract}
Multivalent particles, i.e. microscopic constructs having multiple ligands, can be used to target surfaces selectively depending on their receptor density.  Typically, there is a sharp onset of multivalent binding as the receptor density exceeds a given threshold. However, the opposite case, selectively binding to surfaces with a receptor density below a given threshold, is much harder. Here, we present a simple strategy for selectively targeting a surface with a \emph{low} density of receptors, within a system also having a surface with a higher density of the same receptors. Our strategy exploits competitive adsorption of two species. The first species, called ``guards'', are receptor-sized monovalent particles designed to occupy the high-density surface at equilibrium, while the second multivalent ``attacker'' species outcompetes the guards for binding onto the low-density surface. Surprisingly, the recipe for attackers and guards yields more selective binding with stronger ligand-receptor association constants, in contrast to standard multivalency. We derive explicit expressions for the attacker and guard molecular design parameters and concentrations, optimised within bounds of what is experimentally accessible, thereby facilitating implementation of the proposed approach.
\end{abstract}

\maketitle

\section{Introduction}

Multivalency is a microscopic design strategy for targeting particles with two or more binding units (``ligands'') to a target, such as a surface, having complementary binding units (``receptors'').\cite{Bell:1978hj,Bell:1984wv,Sulzer:1996dd,Huskens:2004je,MartinezVeracoechea:2011kn,Varilly:2012gl,AngiolettiUberti:2013iu,MartinezVeracoechea:2013ih} Nature has exploited multivalency to define interaction paradigms at and between cell surfaces \cite{Bell:1978hj,Macken:1982ia,Bell:1984wv,Perelson:1980ds,Sulzer:1996dd,Sulzer:1997gq,Hlavacek:2002dv,Chen:2003fe,Hong:2007ek,Carlson:2007hv,Shimobayashi:2015fz,Xu:2016kk,Weikl:2016bj,Curk:2017cj,Amjad:2017io,AngiolettiUberti:2017iv,DiMichele:2018dm}, and in the design of bacteria, viruses, and biomolecules themselves. \cite{Mammen:1998im,Kiessling:2013ch,Varner:2015dh,Liese:2018eu} As a result, a large body of research to date has been dedicated to targeting surfaces of cells, cancerous tumours, and other microscopic objects via bio-inspired multivalent interactions. \cite{Shao:2017jd,Wilhelm:2016hh,Ren:2012km,Dubacheva:2015hca,Licata:2008kp,Mahon:2014dn,delaRica:2011bt,Tito:2014kr,Liese:2018eu,Varner:2015dh,Carlson:2007hv,Vonnemann:2015im,Myers:2016ca,Ren:2013fq,Mura:2013fq,Dubacheva:2014ka,Caplan:2005dh,Robinson:2008en,Zitvogel:2014iu,Hong:2007ek,Roh:2015ff,McKenzie:2018fa} Multivalency is also employed to design eloquent self-assembly pathways for synthetic ligand-coated nano- and colloidal particles, often utilising DNA as their binding moities due to their tunable hybridisation free energy. \cite{Mirkin:1996em,Biancaniello:2005ie,Rogers:2011jp,Varilly:2012gl,AngiolettiUberti:2012gi,AngiolettiUberti:2013iu,Wu:2013jg,Stoffelen:2014gr,MejiaAriza:2014jm,AngiolettiUberti:2014kl,Li:2015bm,Wang:2015ep,Curk:2018jb,Srinivasan:2013ez,Grindy:2016jf,Bachmann:2016bb,Newton:2015dq,Theodorakis:2015it,Myers:2016ca,vanderMeulen:2015jb,Newton:2017fp,AngiolettiUberti:2016dd,Mbanga:2016ej,Stoffelen:2015er,DiMichele:2016iu,Zhang:2017kw,Halverson:2016cq} Due to the fact that multiple ligand-receptor bonds are involved in multivalent interactions, their binding kinetics are non-trivial and can complicate the road to reaching equilibrium. \cite{Weikl:2016bj,Bachmann:2016dm,Vijaykumar:2018jq,Licata:2008kp,Newton:2015dq,Newton:2017fp}

The binding affinity of a multivalent particle depends strongly on the number of ligands it has, and the receptor density of the target surface.\cite{MartinezVeracoechea:2011kn, Varilly:2012gl, Tito:2016dk} This is because the binding free energy between the two entities contains a non-trivial entropy term, whose magnitude depends on the number of ligands and receptors. One result of this is superselectivity, where the logarithm of the number of surface-bound particles increases super-linearly with the log of the surface receptor concentration.\cite{MartinezVeracoechea:2011kn} The selectivity becomes larger for particles with more ligands, and when the per-ligand binding energy becomes smaller. Therefore, high-valence particles with weak-binding ligands exhibit sharper surface binding transitions than low-valence particles with strong-binding ligands. This can be used to design multivalent particles that strongly bind to surfaces with many receptors, while having little affinity for surfaces with even a slightly lower density of the same receptors.

A single species of multivalent particles cannot address the opposite scenario, namely targeting a low-receptor-density surface but not one with a higher receptor density. This is because the entropy of binding---the contribution arising from ligand-receptor bonding permutations---always becomes more favorable for a multivalent particle as the surface receptor density increases. Therefore, particles that bind to a surface with few receptors will necessarily bind to one with many. To selectively target only a low receptor density surface, a different approach is needed.

By separately tuning the entropy and energy of binding, mixtures of different kinds of multivalent particles can exhibit ``switch-like'' surface binding.\cite{Tito:2016hh} For example, an equimolar mixture of low-valence nanoparticles with strong-binding ligands can compete with a high-valence weak-ligand species. Both nanoparticle species have the same core size, and exclude the same area when bound to the surface. When the surface receptor density is low, the low-valence species selectively binds to the surface. Upon increasing the receptor concentration, there is a switch-point, after which the surface becomes occupied by the high-valence species. The surface receptor density thus acts to shift the balance between the entropic and energetic terms in the free energy of binding for the two species. The binding free energy of the low-valence strong-binding species is dominated by the energetic term; on the other hand, the high-valence weak-binding species has a substantial entropy of binding.

The present work takes this as inspiration, and devises an ``attacker and guard'' strategy for selectively targeting a surface with low receptor density within a system that also has a surface with a higher density of the same receptors. This might be, for example, two populations of cells in a suspension, with one population having a high membrane concentration of a particular receptor, and the other having a low concentration of the same receptors.

The strategy we propose entails using one species of particles, called ``guards'', to occupy the receptors on the high-density surface. These particles have a size on the order of a single receptor. A second larger species of particles, called ``attackers'', are then designed to out-compete the guards for binding on the low-receptor-density surface at equilibrium, but not on the high-density surface. Experimental accessibility and robustness are emphasised in devising this recipe. The strategy may prove useful for selectively imaging cell surfaces \emph{in vitro} that have globally or locally low receptor density, e.g. by making the attackers fluorescently active and the guards not via a DNA-PAINT approach. \cite{Delcanale:2018fy} This approach may also have use in selective sequestration or aggregation of microscopic entities with a low receptor density, in which the attackers act as the aggregating agents.

\section{Tuning multivalent binding by microscopic construction}

\begin{figure*}
	\centering
	\includegraphics[width= 0.90\textwidth]{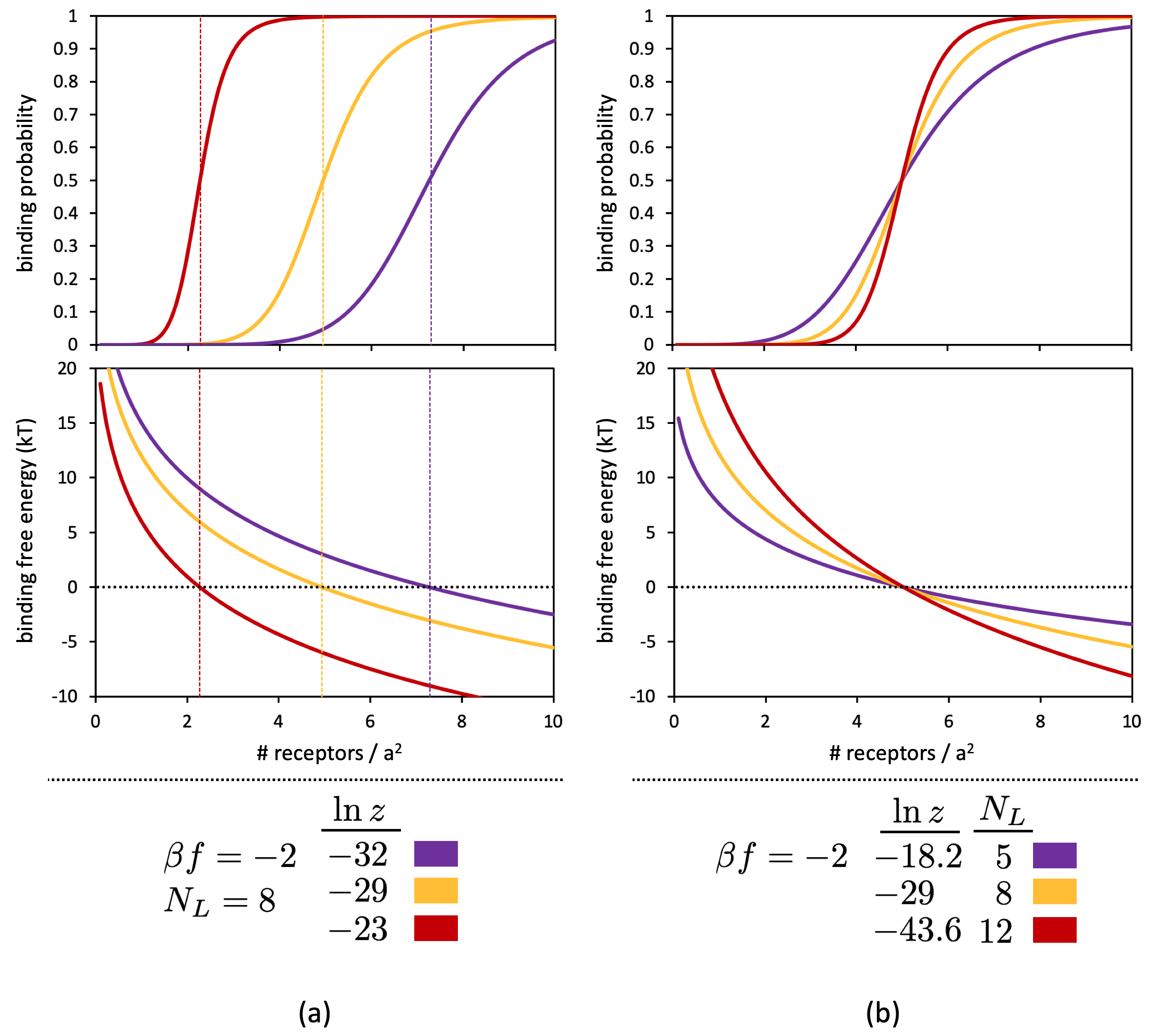}
	\caption{ Multivalent binding probability (Eq. \ref{eqn:MVBindingP}) (upper panels) and binding free energy (lower panels) as a function of the number of receptors $N_R$ per surface lattice site of size $a^2$. In (a), the ligand-receptor binding free energy $\beta f = -2$ and number of ligands $N_L = 8$ are kept fixed, while the fugacity of the particles is set to $\ln{z} = -32, -29, -23$ (purple, yellow, red). Vertical dashed lines indicate inflection points for each adsorption profile. In (b), the particle ligand-receptor binding free energy is also fixed at $\beta f = -2$, while the number of ligands on the particle is set to $N_L = 5, 8, 12$ (purple, yellow, red). In each case, the fugacity is adjusted such that the binding transition occurs at $N_R = 5$.}
	\label{fig:MVBindingVsNR}
\end{figure*}  

To begin, we briefly review how the binding free energy of a multivalent particle dictates its affinity to binding to a surface, as a function of the target surface's receptor density. This is important for understanding how to manipulate the binding affinity of two competing species for the more complex case of targeting a low-receptor-density surface.

The binding free energy of a multivalent particle is given by the standard relation $\beta G(N_R) = -\ln{Q(N_R)} - \ln{z}$, where $Q$ is the partition function for the particle when it is adjacent to the receptor surface, and $z$ is the fugacity of the particles in solution above the receptor surface. The quantity $N_R$ is the number of receptors that are accessible to the particle when it is adjacent to the surface. This is defined as $N_R = \sigma_R a^2$, where $\sigma_R$ is the number of receptors per unit area on the surface (the ``receptor density''), and $a$ is the diameter of the multivalent particle and its ligands. We refer to one $a^2$-sized area element of the receptor surface as a surface ``lattice site''.

 Multivalent particles at low concentration can be assumed to follow ideal gas statistics, in which the fugacity $z$ is related to the molar concentration ``$[C]$'' of the particles in solution by 
\begin{equation}
	[C] = \frac{z}{N_A v_{ex}}.
	\label{eqn:EqForConcentrationFromFug}
\end{equation}
Here, $N_A$ is Avogadro's number, and $v_{ex}$ is the excluded volume (or ``localisation volume'') for one multivalent particle. Note that $v_{ex}$ must be in units of decimeters$^3$ in order for the concentration $[C]$ to be in the appropriate units of moles / litre.

When a multivalent particle is adjacent to a binding surface, it can have one or more additional ``non-specific'' (i.e. non-multivalent) interaction free energy contributions. Collecting all of these additional contributions into the quantity ``$G_{NS}$'', then the total binding free energy of the multivalent particle is
\begin{equation}
	\beta G(N_R) = -\ln{Q(N_R)} - \ln{z} + \beta G_{NS}.
	\label{eqn:MVBindingFreeEnergyAllTerms}
\end{equation}	
Written this way, \emph{the partition function $Q(N_R)$ explicitly contains only the ``multivalent'' ligand/receptor bonding contributions to the binding free energy}. Unless otherwise noted, we henceforth set $\beta G_{NS}$ to zero, and focus attention on the multivalent binding contributions in $Q(N_R)$.

When receptors are immobile and \emph{uniformly} placed on the surface at a density of $\sigma_R$, the partition function $Q(N_R)$ for a lattice site when occupied by a multivalent particle  is well described\cite{MartinezVeracoechea:2011kn,Curk:2017cj} by
\begin{equation}
	Q(N_R) = \sum_{\lambda = 0}^{N^*}{\binom{N_R}{\lambda} \binom{N_L}{\lambda} \lambda! e^{-\beta \lambda f}}
	\label{eqn:MVPartitionFunctionExact}
\end{equation}
where $f$ is the free energy for forming a single ligand-receptor bond, $N_L$ is the effective number of ligands on the multivalent particle that can access the surface for receptor binding at any given time, and $N^* = \min{(N_R, N_L)}$. This expression effectively treats the receptors and ligands as an ideal gas within the surface lattice site, while crucially enforcing that they each may only have zero or one binding partner in any given microstate.

Note that the quantity $N_L$ is almost always less than the \emph{total} number of ligands on the particle. This is because not all ligands can simultaneously reach the receptor surface for binding, depending on how the multivalent particle is oriented relative to the surface. As an example, for a large colloidal particle with short ligands, a simple way to estimate $N_L$ is to multiply the average grafting density $\sigma_L$ of ligands on the colloid by the average contact area $A_{\text{contact}}$ between the colloid and the receptor surface. This ``effective valence'' $A_{\text{contact}} \sigma_L$ is much smaller than the ``total'' valence $A_{\text{total}} \sigma_L$ of the colloid, where $A_{\text{total}}$ is the colloid's total surface area. For multivalent constructs, it is the ``effective'' valence that dictates the binding behaviour, and this is what the quantity $N_L$ signifies throughout our discussion. 

Adding in Poisson fluctuations to the number of receptors in each lattice site considerably simplifies this expression. With a Poisson distribution centered around a mean value of $N_R$ receptors per lattice site, the probability the multivalent particle ``sees'' $j$ receptors within area $a^2$ follows $P(j; N_R) = e^{-N_R} N_R^j / j!$. In this case, Eq. \ref{eqn:MVPartitionFunctionExact} simplifies to\cite{Curk:2017cj} 
\begin{equation}
	Q(N_R) = \left(1 + N_R e^{-\beta f}\right)^{N_L}.
	\label{eqn:MVPartitionFunction}
\end{equation}
Strictly speaking, this form is only exact  when the coverage of bound multivalent particles on the receptor surface is low, such that each bound particle can independently sample the Poisson distribution of receptors. For higher surface coverage, the adsorption statistics become multi-Langmuir, and this must be calculated numerically (see Ref. \citenum{Dubacheva:2019gq} for a mathematical and experimental discussion of this regime). To proceed analytically for the present discussion, we adopt Eq. \ref{eqn:MVPartitionFunction} and then remark on potential discrepancies compared to the more exact multi-Langmuir adsorption later on in the discussion.  Note that Eq. \ref{eqn:MVPartitionFunction} is also exact when the receptors are mobile on the surface, non-depletable (i.e. coming from a grand canonical reservoir), and at an average concentration of $N_R$ per lattice site. However, for the present study we assume that the receptors are immobile over the timescale of multivalent particle binding and equilibration. 

Given $Q(N_R)$ in Eq. \ref{eqn:MVPartitionFunction}, the binding free energy for a multivalent particle is
\begin{equation}
	\beta G(N_R) = -N_L \ln{\left(1 + N_R e^{-\beta f}\right)} - \ln{z} + \beta G_{NS}.
	\label{eqn:MVBindingFE}
\end{equation}
The probability that a surface lattice site is occupied by a multivalent particle is then
\begin{equation}
	P_b(N_R) = \frac{e^{-\beta G(N_R)}}{1 + e^{-\beta G(N_R)}}.
	\label{eqn:MVBindingP}
\end{equation}
When $\beta G(N_R)$ is greater than zero (i.e. an unfavourable binding free energy change), then $P_b(N_R)$ goes to zero. Similarly, when $\beta G(N_R)$ is less than zero, corresponding to a favourable free energy of binding, then $P_b(N_R)$ goes to unity. Thus, $\beta G(N_R) = 0$ corresponds to the binding transition, and the derivative of $\beta G(N_R)$ with respect to $N_R$ at $\beta G(N_R) = 0$ reflects the sharpness of the transition.

Equation \ref{eqn:MVBindingFE} shows us how the multivalent binding free energy changes with $N_R$, $f$, $N_L$, $z$, and $\beta G_{NS}$. The fugacity $z$ ($\propto$ concentration $[C]$) shifts the binding free energy $\beta G(N_R)$ up or down by a constant. It therefore provides a convenient handle for adjusting the receptor density $\sigma_R$ at which the adsorption transition occurs. This is illustrated in Figure \ref{fig:MVBindingVsNR}a. As the fugacity grows smaller, then the overall binding free energy shifts higher (more unfavourable). The receptor density $\sigma_R$ where the adsorption transition occurs correspondingly increases, and the sharpness of the transition decreases (as the local derivative of $\beta G(N_R)$ for increasing $N_R$ gets smaller).

The sharpness of the binding transition can be tuned by adjusting the molecular construction of the multivalent particle, via its valence $N_L$ and ligand-receptor binding strength $f$. Figure \ref{fig:MVBindingVsNR}b shows examples of tuning the adsorption sharpness by changing $N_L$. In each case, the fugacity has been tuned so that the adsorption transition is centered at $N_R = 5$. Making $N_L$ larger causes the gradient of $\beta G(N_R)$ with $N_R$ to be steeper and more negative, leading to a sharper binding transition.

 The free energy curves in Figures \ref{fig:MVBindingVsNR}a and b can also be vertically shifted by altering the ``non-specific'' binding free energy $\beta G_{NS}$. The quoted values of $\ln{z}$ in those examples can, for example, be equivalently interpreted as ``effective'' fugacities given by $\ln{z_{\text{true}}} - \beta G_{NS}$, where the former is the true solution fugacity determined by the multivalent particle concentration. Since $\beta G_{NS}$ can also be tuned by chemical design in principle, then it is an additional adjustment knob for \emph{uniformly} shifting the multivalent binding free energy if tuning the particle concentration proves to be impractical. (An example would be a particular target receptor surface that requires a vanishingly small or infeasibly large bulk solution concentration of multivalent particles in order to reach a desired binding equilibrium).

\section{Targeting a low-receptor-density surface with two competing species}

\begin{figure*}
	\centering
	\includegraphics[width= 0.90\textwidth]{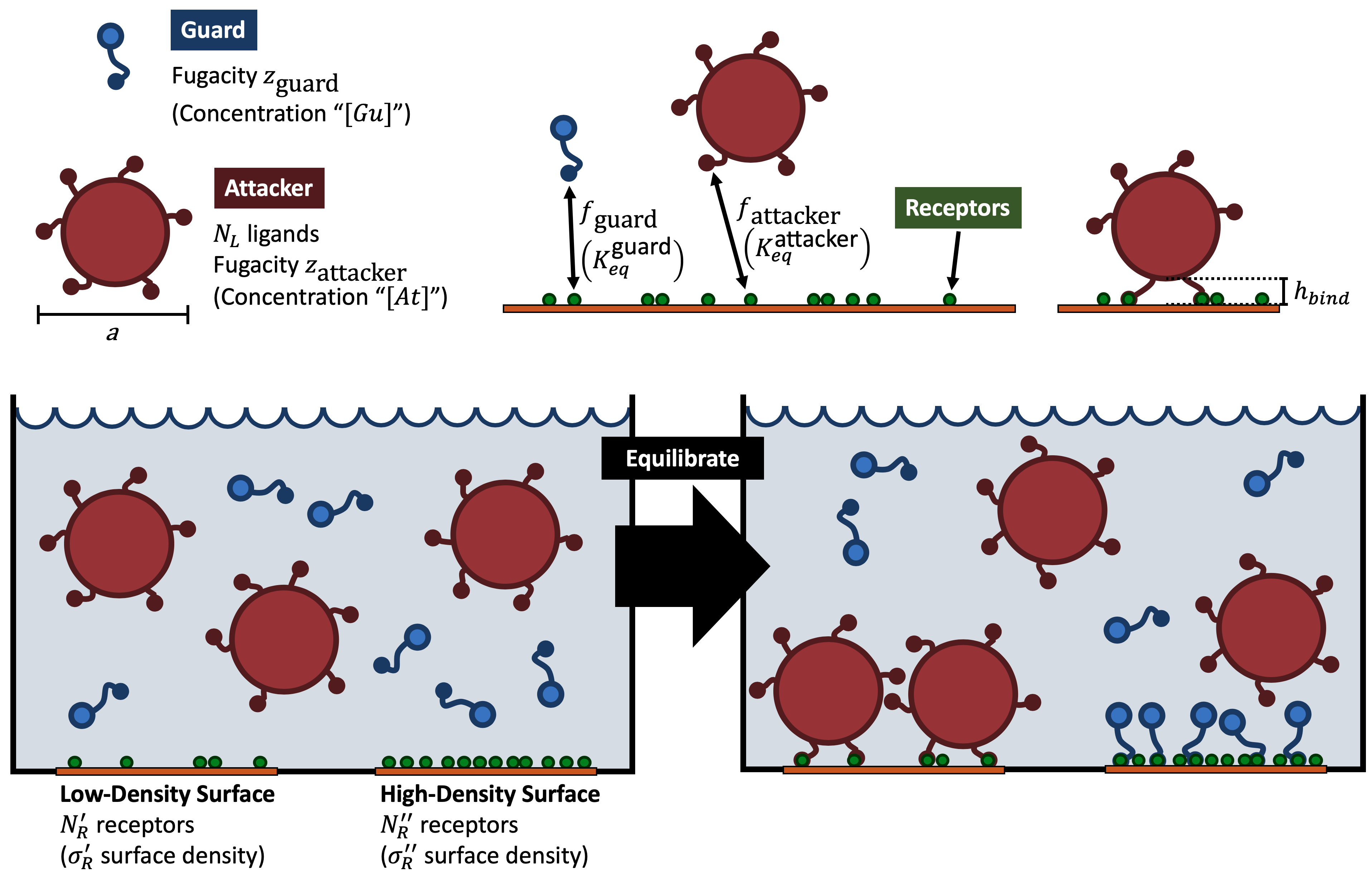}
	\caption{ Graphical depiction of the ``Attacker $\&$ Guard'' targeting strategy, showing key ingredients, mathematical parameters, and sought-after equilibrium binding distribution of particles. Upper half of image shows attackers, guards, and a receptor surface, along with relevant mathematical parameters (described in main text). Lower half of image shows attackers and guards in a hypothetical solution containing a low- and high-receptor-density surface. On the left, attackers and guards have just been added, while on the right, equilibrium has been reached. }
	\label{fig:OverallCartoon}
\end{figure*}  

The binding free energy of a multivalent particle, $\beta G(N_R)$, can be completely tailored by the parameters $f$, $N_L$, and $z$. This can be used to design a multivalent species that binds strongly to a surface with a high receptor density, while not binding to one with a lower density. However, it is impossible to achieve the opposite case with just one multivalent binder. As is apparent in Figures \ref{fig:MVBindingVsNR}a and b, the binding free energy cannot be manipulated in such a way that the multivalent particle only binds at low surface receptor concentration.

To solve this problem, we can introduce a second particle species that competes for surface binding with the original species. The goal is to define the second species such that it blocks the first from binding to the high-receptor-density surface, but not the low-density surface. We call this second species the ``guards'', and the original species the ``attackers''.  The forthcoming ``Attacker $\&$ Guard'' strategy is graphically depicted in Figure \ref{fig:OverallCartoon}, showing the molecular ingredients, salient mathematical parameters, and intended equilibrium distribution of attackers and guards on the two receptor surfaces. 

To start, the guards are defined to be a roughly \emph{receptor-sized monovalent} species. The free energy of binding for a monovalent species is \emph{independent} of the surface receptor density. Each receptor has a partition function of the form
\begin{equation}
	q_{\text{receptor}} = 1 + z_{\text{guard}} e^{-\beta f_{\text{guard}}},
\end{equation}
where $f_{\text{guard}}$ is the free energy for forming a guard-receptor bond, and $z_{\text{guard}}$ is the guard fugacity. The first term in $q_{\text{receptor}}$ is the weight for when the receptor is not bound to anything, and the second is for when it is bound to a guard particle. The free energy of a single receptor is then just $-\ln{(q_{\text{receptor}})}$.

However, to compare the guard binding free energy to the attackers, we must consider the total free energy of guard binding over the full area $a^2$ occupied by an attacker. This quantity depends \emph{linearly} on the number $N_R$ of receptors within $a^2$:

\begin{align}
	\beta G_{\text{guard}}(N_R) &= -\ln{\left(q_{\text{receptor}}^{N_R}\right)} \nonumber \\
	&= -N_R \ln{\left(1 + z_{\text{guard}} e^{-\beta f_{\text{guard}}}\right)} \nonumber \\
	&= -N_R C_{\text{guard}}.
	\label{eqn:FEGuards}
\end{align}
Here, the combined tunable guard parameter
\begin{equation}
	C_{\text{guard}} \equiv \ln{\left(1 + z_{\text{guard}} e^{-\beta f_{\text{guard}}}\right)}
	\label{eqn:GuardParameter}
\end{equation}
has been defined for notational clarity in subsequent equations.

On the other hand, the free energy of the lattice site when occupied by an attacker depends \emph{logarithmically} on $N_R$ in the lattice site, via Eq. \ref{eqn:MVBindingFE}:
\begin{align}
	\beta G_{\text{attacker}}(N_R) = &-N_L \ln{\left(1 + N_R e^{-\beta f_{\text{attacker}}}\right)} - \ln{z_{\text{attacker}}}.
	\label{eqn:FEAttackers}
\end{align}
assuming no non-specific binding free energy contribution $\beta G_{NS}$.
Importantly, we assume that \emph{when an attacker is bound, it excludes all receptors over the area $a^2$ from binding to any guards}. This assumption is most likely to hold when the attacker is a solid structure like, e.g., a ligand-coated nanoparticle, vesicle, or virus. The implications of this assumption breaking down are examined later.

\begin{figure}
	\centering
	\subfigure[]{\includegraphics[width= 0.49\textwidth]{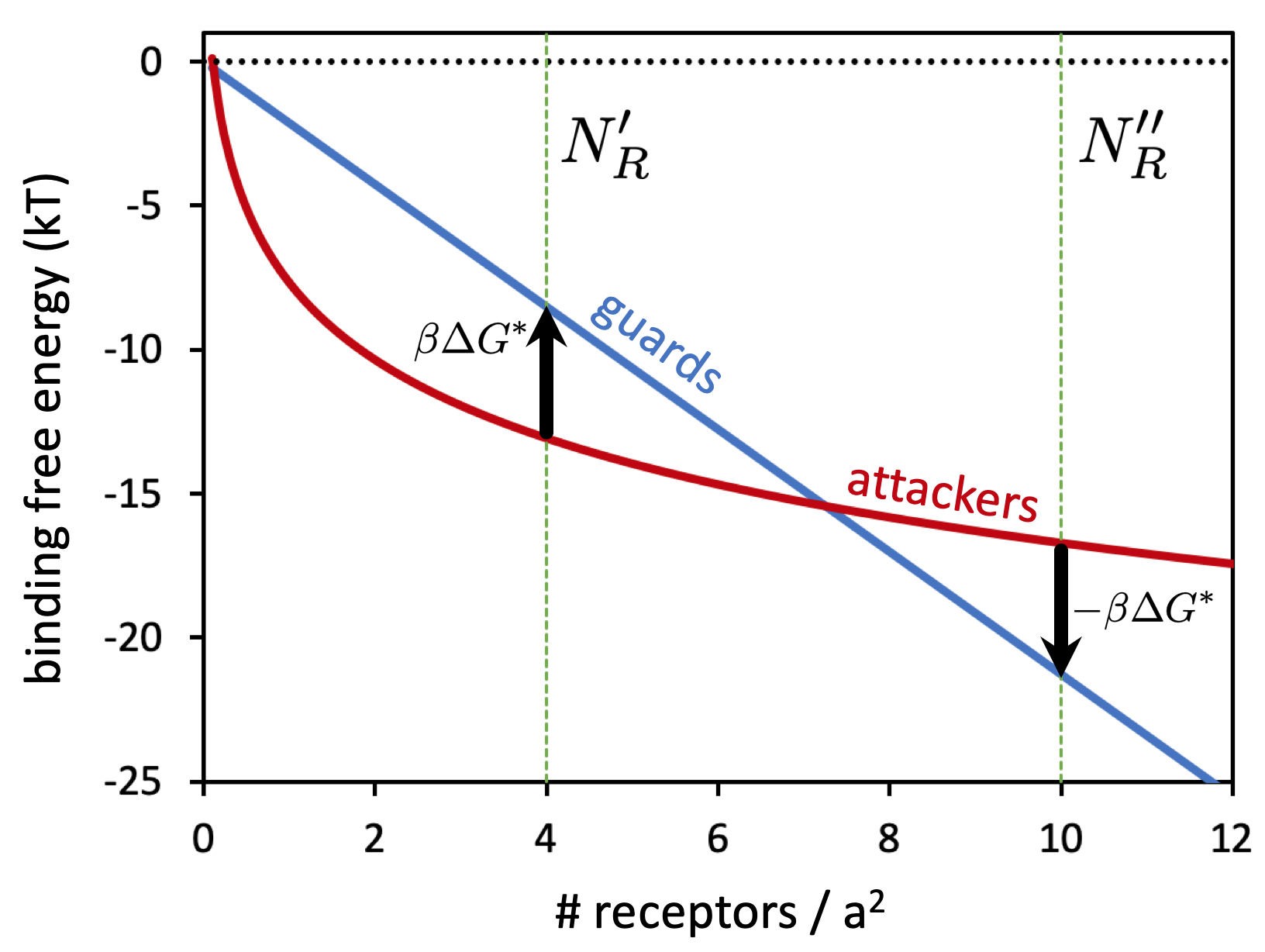}}
	\subfigure[]{\includegraphics[width= 0.49\textwidth]{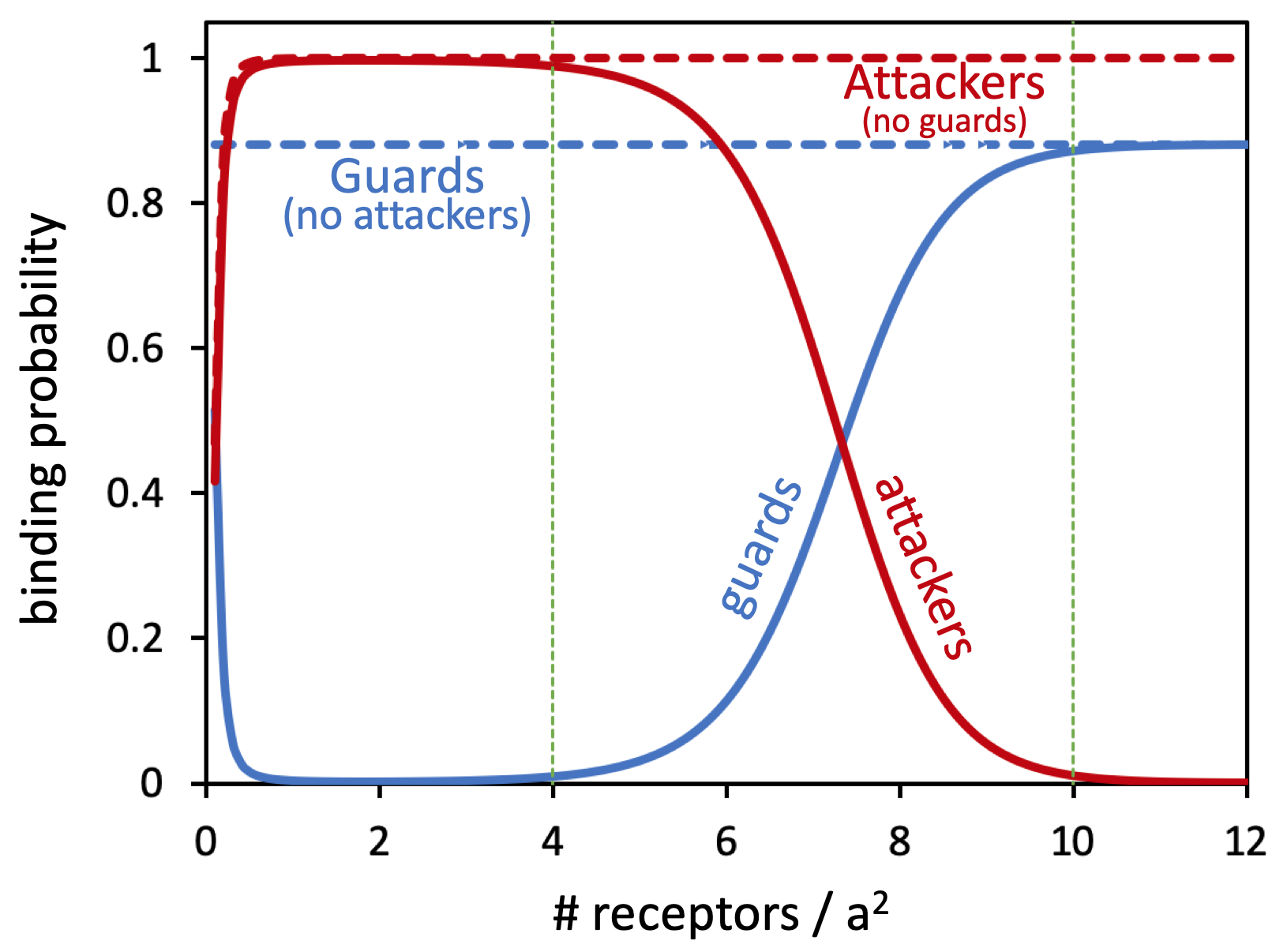}}
	\caption{ Binding free energy (a, via Eqs. \ref{eqn:FEGuards} and \ref{eqn:FEAttackers}) and surface adsorption probability (b, via Eqs. \ref{eqn:AttackerBindingProb} and \ref{eqn:GuardBindingProb}) for attackers (red curves) and guards (blue curves) as a function of the average number of receptors per surface lattice site of area $a^2$. Green vertical lines indicate the low-receptor-density and high-receptor-density surfaces, with $N_R' = 4$ and $N_R'' = 10$, respectively. In (b), solid lines are for when attackers and guards coexist in the same system, while dashed lines are for when they are separately in the system. Multivalent attacker parameters are $N_L = 4$, $\beta f_{\text{attacker}} = -3$, and $\ln{z_{\text{attacker}}} = -4.53$, while monovalent guard parameters are $\beta f_{\text{guard}} = -4$ and $\ln{z_{\text{guard}}} = -2$. Attacker fugacity $z_{\text{attacker}}$ has been chosen such that $\beta \Delta G(N_R') = - \beta \Delta G (N_R'')$, indicated as $\beta \Delta G^*$ in (a) here. }
	\label{fig:AttackerGuardBinding}
\end{figure}  

The different scaling of the attacker and guard binding free energies (per lattice site) with $N_R$ can be exploited as shown in Figure \ref{fig:AttackerGuardBinding}a. In this example, we suppose that the low-density surface has $N_R = 4 \equiv N_R'$, and the high-density surface has $N_R = 10 \equiv N_R''$. The blue curve in Figure \ref{fig:AttackerGuardBinding}a plots $\beta G_{\text{guard}}(N_R)$ as a function of $N_R$. The slope of the line is controlled by the guard fugacity $z_{\text{guard}}$ and binding strength $f_{\text{guard}}$. The red curve in Figure \ref{fig:AttackerGuardBinding}a displays $\beta G_{\text{attacker}}(N_R)$. The weaker logarithmic dependence of $\beta G_{\text{attacker}}(N_R)$ on $N_R$ has been exploited to tune the attacker's design (via $N_L$, $f_{\text{attacker}}$, $z_{\text{attacker}}$) such that: at $N_R'$, $\beta G_{\text{attacker}}(N_R') < \beta G_{\text{guard}}(N_R')$; while at $N_R''$, $\beta G_{\text{guard}}(N_R'') < \beta G_{\text{attacker}}(N_R'')$.

The resulting binding behaviour of the attackers and guards is displayed in Figure \ref{fig:AttackerGuardBinding}b. The probability that a surface site is occupied by an attacker is
\begin{equation}
	P_b^{\text{attacker}} (N_R) = \frac{e^{-\beta \Delta G(N_R)}}{1 + e^{-\beta \Delta G(N_R)}},
	\label{eqn:AttackerBindingProb}
\end{equation}
where
\begin{equation}
	\beta \Delta G(N_R) \equiv \beta G_{\text{attacker}}(N_R) - \beta G_{\text{guard}}(N_R).
\end{equation}
For comparison, we also plot the probability that a single receptor is attached to a guard:
\begin{align}
	P_b^{\text{guard}}(N_R) = &\left( \frac{z_{\text{guard}} e^{-\beta f_{\text{guard}}}}{1 + z_{\text{guard}} e^{-\beta f_{\text{guard}}}}\right) \nonumber \\
	&\times \left(1 - P_b^{\text{attacker}} (N_R)\right).
	\label{eqn:GuardBindingProb}
\end{align}
These are derived in Appendix \ref{app:MonovalentGuardBinding}.

In Figure \ref{fig:AttackerGuardBinding}, the guard and attacker fugacities have been set to $\ln{z_{\text{guard}}} = -2$ and $\ln{z_{\text{attacker}}} = -4.53$. Using Eq. \ref{eqn:EqForConcentrationFromFug}, a rough feasibility check can be made for the molar concentrations these fugacities correspond to assuming that there is no non-specific binding free energy ($\beta G_{NS}$) contribution. For example, considering ``receptor-sized'' guards of length $10$nm ($\approx 1$ hemagglutinin unit on the exterior of an influenza virus particle), and attackers of size $100$nm, then these fugacities correspond to guard and attacker concentrations of around $[Gu] \approx 200$ $\mu$M, and $[At] \approx 18$ nM. These concentrations, along with the choices of $\beta f_{\text{guard}} = -4$, $N_L = 4$, and $\beta f_{\text{attacker}} = -3$ in that figure, are well within accessible experimental range.

At $N_R'$, the attackers are strongly bound, while at $N_R''$, the guards outcompete the attackers for binding. The value of $N_R$ where $\beta G_{\text{guard}}(N_R) = \beta G_{\text{attacker}}(N_R)$ defines the ``switch point'' between the two species. Combining both species into the same system is essential, as alone, one or the other species would strongly bind to both the low- and high-density surfaces (dashed blue and red lines in Figure \ref{fig:AttackerGuardBinding}a).

The effectiveness of the targeting recipe is assessed by the difference in attacker binding probabilities at $N_R'$ and $N_R''$, defined as
\begin{equation}
	\text{Effectiveness} \equiv \phi \equiv P_b^{\text{attacker}}(N_R') - P_b^{\text{attacker}}(N_R'').
	\label{eqn:Effectiveness}
\end{equation}
Values of $\phi$ near unity are optimal, $0$ means that the attackers bind equally well to $N_R'$ and $N_R''$, while negative values (approaching $-1$) mean that the attackers favour binding to $N_R''$ rather than $N_R'$.

In order to achieve an effectiveness $\phi$ of unity, the attacker/guard binding free energy difference $\beta \Delta G(N_R)$ must go to negative infinity at $N_R'$ and positive infinity at $N_R''$. Therefore, to proceed further, we seek the attacker and guard design parameters that lead to a given \emph{chosen} effectiveness $\phi$.

By inspection of Figure \ref{fig:AttackerGuardBinding}a, the maximum possible effectiveness of an attacker+guard design is set by the slope of the guard free energy (blue line) as a function of $N_R$. The larger and more negative the slope, the larger the free energy difference between $N_R'$ and $N_R''$. The connection between a desired $\phi$, and the necessary $C_{\text{guard}}$, is developed in the next section.

The remaining task is to find an optimal attacker design, i.e. $N_L$, $z_{\text{attacker}}$, and $f_{\text{attacker}}$. The best attacker design is one in which their free energy of binding (red curve) as a function of $N_R$ is nearly constant between $N_R'$ and $N_R''$, so as to bisect the blue guard curve in this interval in Figure \ref{fig:AttackerGuardBinding}a. Doing so obtains the most negative $\beta \Delta G(N_R')$ and most positive $\beta \Delta G(N_R'')$. Based on Eq. \ref{eqn:FEAttackers}, this occurs when the attacker has \emph{few strong-binding} ligands.

The number of ligands $N_L$ and ligand/receptor bonding strength $f_{\text{attacker}}$ of the attackers are set by chemical design. Supposing these are pre-defined for the time being, we then seek the attacker fugacity (concentration) $z_{\text{attacker}}$ that maximises the targeting effectiveness $\phi$. Attackers can be readily titrated into a system so that the level of control over $z_{\text{attacker}}$ is very high compared to the molecular design. It is therefore a convenient experimental control parameter to optimise over for the attackers.

The attacker fugacity $z_{\text{attacker}}$ that maximises the targeting effectiveness is where
\begin{equation}
	\frac{d \phi}{d \ln{z_{\text{attacker}}}} = 0.
\end{equation}
This is carried out in Appendix \ref{app:SelectivityAndTolerance}, where we find that the largest effectiveness is obtained by choosing $z_{\text{attacker}}$ such that
\begin{equation}
	\beta \Delta G(N_R') = -\beta \Delta G(N_R'').
	\label{eqn:OptimalCondition}
\end{equation}
We call this optimum ``$\beta \Delta G^*$'': a free energy ``gap'' that is directly tuned by the design of the guards and attackers. Inserting the condition in Eq. \ref{eqn:OptimalCondition} into Eq. \ref{eqn:Effectiveness} yields
\begin{equation}
	\text{Optimal Effectiveness} \equiv \phi^* = \frac{e^{-\beta \Delta G^*} - 1}{e^{-\beta \Delta G^*} + 1}.
	\label{eqn:OptSelectivity}
\end{equation}
Tuning of the attacker fugacity to satisfy this optimum condition has been carried out in Figure \ref{fig:AttackerGuardBinding}a, and the symmetric free energy gap $\beta \Delta G^*$ is indicated. The ``tolerance'' of the targeting design is quantified by the width of the optimum in targeting effectiveness at $\beta \Delta G(N_R') = -\beta \Delta G(N_R'') = \beta \Delta G^*$: 
\begin{align}
	\text{Design Tolerance for }\phi^* &\equiv -\left(\frac{d^2 \phi}{\left[d \ln{z_{\text{attacker}}}\right]^2}\right)^{-1} \nonumber \\
	&\approx \frac{1}{2} e^{-\beta \Delta G^*}.
	\label{eqn:OptTolerance}
\end{align}

Equations \ref{eqn:OptSelectivity} and \ref{eqn:OptTolerance} indicate that \emph{the most effective and most tolerant targeting design is obtained when the free energy gap $\beta \Delta G^*$ is large and negative}.

\section{Optimal Design of Attackers and Guards}

The quantity $\beta \Delta G^*$ is directly tuned by the molecular design of the attackers and guards. The necessary $\beta \Delta G^*$ in order to achieve a given effectiveness $\phi^*$ is obtained by inverting Eq. \ref{eqn:OptSelectivity}, yielding $\beta \Delta G^*_{\text{needed}} = -\ln{\left(\frac{1 + \phi^*}{1 - \phi^*}\right)}$. The free energy gap in terms of the attacker and guard binding free energies is $\beta \Delta G^* = \beta G_{\text{attacker}}(N_R') - \beta G_{\text{guard}}(N_R')$, subject to the constraint $\beta G_{\text{attacker}}(N_R') - \beta G_{\text{guard}}(N_R') = \beta G_{\text{guard}}(N_R'') - \beta G_{\text{attacker}}(N_R'')$ given by Eq. \ref{eqn:OptimalCondition}.

These three relations, applied to the guard and attacker binding free energies (Eqs. \ref{eqn:FEGuards} and \ref{eqn:FEAttackers}), result in closed-form expressions for the necessary attacker and guard solution concentrations $[At]^*$ and $[Gu]^*$ to achieve a desired targeting effectiveness $\phi^*$. The derivation is carried out in Appendix \ref{app:OptimalDesign}, resulting in
\begin{widetext}
\begin{align}
	[At]^*  = &\frac{1}{N_A h_{bind} a^2}  \dfrac{\left[\left(\dfrac{q_L(c_R'')}{q_L(c_R')}\right)^{\frac{N_L}{2}} \left(\dfrac{1 + \phi^*}{1 - \phi^*}\right)\right]^{\frac{c_R'' + c_R'}{c_R'' - c_R'}}}{\left[q_L(c_R') q_L(c_R'')\right]^{\frac{N_L}{2}}};
	\label{eqn:AtDesign}
\end{align}
\begin{equation}
	[Gu]^*  = \frac{1}{K_{eq}^{\text{guard}}} \left\{\left[\left(\frac{q_L(c_R'')}{q_L(c_R')}\right)^{\frac{N_L}{2}} \left(\frac{1 + \phi^*}{1 - \phi^*}\right)\right]^{\frac{2}{N_A a^2 \left(c_R'' - c_R'\right)}} - 1\right\}.
	\label{eqn:GuDesign}
\end{equation}
\end{widetext}
where $q_L(c_R')$ and $q_L(c_R'')$ are dimensionless quantities calculated by
\begin{equation}
	q_L(c_R) = \left(1 + \frac{c_R K_{eq}^{\text{attacker}}}{h_{bind}}\right).
\end{equation}
The expressions are presented here in chemical equilibrium form, to facilitate experimental implementation. Appendix \ref{app:OptimalDesign} details the mathematical transformations involved to translate the statistical mechanical theory into the form shown here.

The molar equilibrium association constants $K_{eq}^{\text{attacker}}$ and $K_{eq}^{\text{guard}}$ are for, respectively: binding between a free receptor and a single free ligand on the attacker; and between a free receptor and a guard particle in solution. The quantities $c_R'$ and $c_R''$ are ``surface receptor molarities'' (in units of moles of receptors per unit surface area) on the low- and high-density surfaces, respectively, and $a$ is the diameter of the attacker (including its ligand corona). These are related to the previously-employed surface densities $\sigma_R$ by $c_R = \sigma_R / N_A$. The quantity $h_{bind}$ is the equilibrium binding height of the attacker; its precise definition depends on the type of construct the attacker is, to be discussed shortly. Finally, $N_L$ is the effective number of ligands on the attacker species that can access the surface for receptor binding at any given time.

Equations \ref{eqn:AtDesign} and \ref{eqn:GuDesign} provide physical insight. As the desired effectiveness $\phi^*$ is increased to unity, then the required attacker and guard concentrations (or association constants) grow very large. Therefore, \emph{the low-receptor density surface can be targeted more effectively when the attackers and guards have a larger overall binding affinity}. This is the opposite to standard one-component multivalent targeting of high-density surfaces, in which weak ligand-receptor bonds yield higher selectivity.

The remainder of this section takes a closer look at the guard and attacker parameters in Eqs. \ref{eqn:AtDesign} and \ref{eqn:GuDesign}. The kinetics involved to reach binding equilibrium between the two competing adsorbers are illustrated, and a suggested ``recipe'' for adding attackers and guards that circumnavigates potential kinetic barriers is outlined. To finish, factors that enhance the tolerance of the design are discussed, while also remarking on effects not considered in our theory which might \emph{reduce} the targeting effectiveness.

\subsection{Equilibrium constants \& attacker binding height in the context of experiment}

 The binding constant $K_{eq}^{\text{attacker}}$ for the attacker ligands includes the enthalpic contribution to the ligand/receptor bond, as well as the extra (entropic) free energy cost $\Delta G_{\text{lig,cnf}}$ for bond formation. Usually in experiment, the ligand/receptor binding constant is only measured for the case where the receptor and ligand structures are free in solution and untethered to any host surfaces. This reference ligand/receptor binding constant, ``$K_{eq}^{\circ,\text{attacker}}$'', is mapped to the binding constant for the ligands when attached to the attacker particle by
\begin{equation}
	K_{eq}^{\text{attacker}} = K_{eq}^{\circ,\text{attacker}} e^{-\beta \Delta G_{\text{lig,cnf}}}.
\end{equation}
Detailed discussions and models for approximating the extra entropic free energy penalty $\Delta G_{\text{lig,cnf}}$ for ligand/receptor binding are given in Refs. \citenum{Varilly:2012gl} and \citenum{MartinezVeracoechea:2013ih}.

\begin{figure}
	\centering
	\includegraphics[width= 0.40\textwidth]{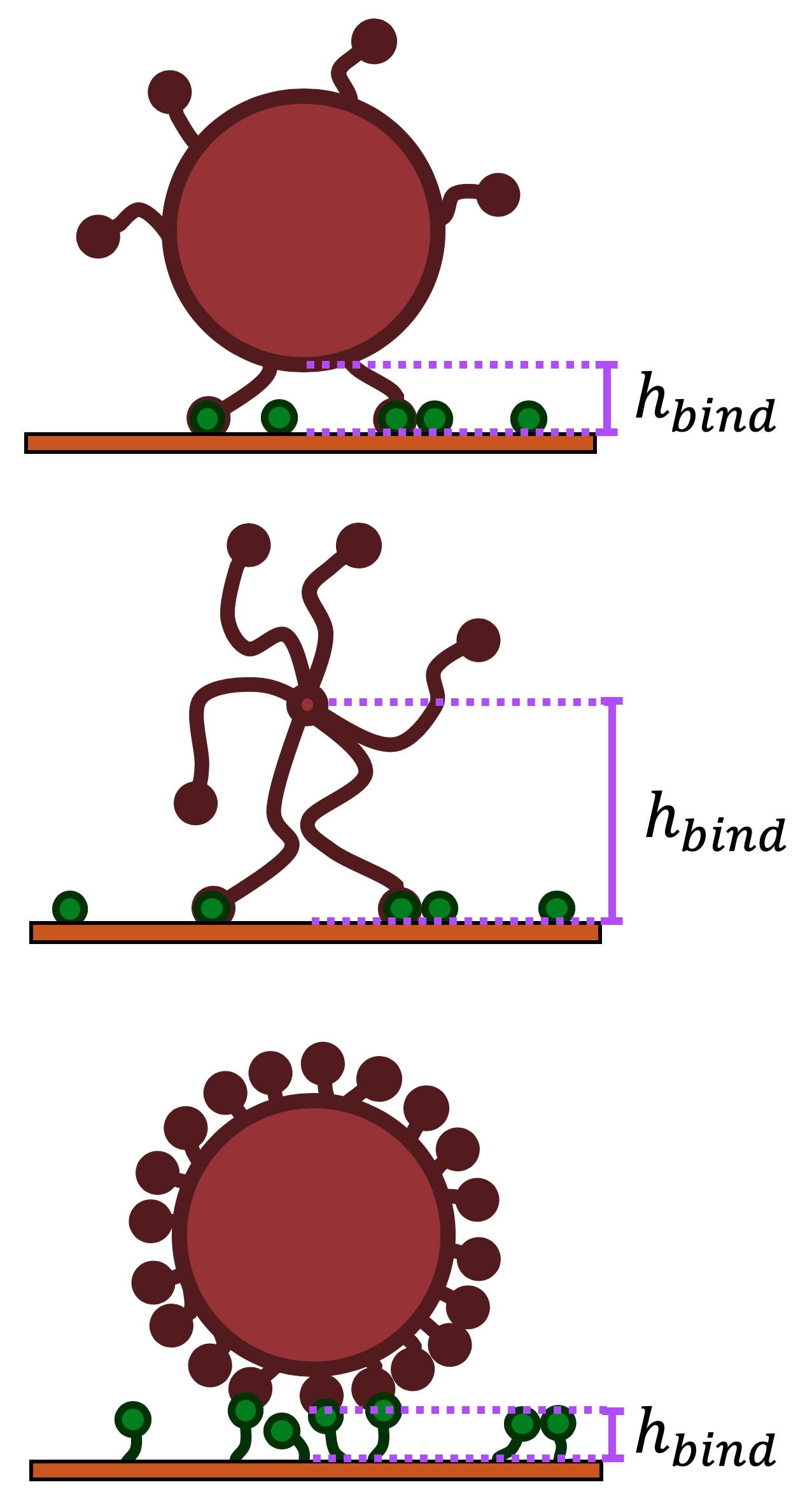}
	\caption{ Possible definitions for the equilibrium binding height parameter $h_{bind}$, depending on the multivalent attacker structure. Top image depicts a multivalent particle with a solid core and flexible ligands, interacting with a surface of short/inflexible receptors. Middle image is a star polymer-like structure or dendrimer-like structure interacting with short/inflexible receptors. Finally, bottom image is a spherical structure with densely-packed short/inflexbile ligands (e.g. like an influenza virus) interacting with flexible receptors.}
	\label{fig:HBinds}
\end{figure}  

The quantity $h_{bind}$ is a binding distance parameter. It  controls the receptor ``effective molarity'' that the ligands on an attacker ``see'' when the attacker is surface-bound.\cite{Curk:2017cj} Appendix \ref{app:OptimalDesign} describes this in greater mathematical detail.

The choice of how to precisely define $h_{bind}$ depends on the type of construct that the attacker is, as illustrated in Figure \ref{fig:HBinds}. For example, if the attacker is a star polymer or dendrimer construct \cite{Licata:2008kp}, then $h_{bind}$ should be taken to be the distance between the receptor surface and the center of the star. If the attacker is a solid particle-type construct like a DNA-coated nanoparticle or a surface-functionalised vesicle, then $h_{bind}$ should be defined as the equilibrium distance between the outer surface of the attacker, and the receptor surface. 

In the latter case, whether or not $h_{bind}$ should include the lengths of the receptors or ligands depends on how densely packed they are on their respective surfaces. For example, if the attacker has a dense packing of binding ligands (like, e.g., the influenza virus), then one could argue that $h_{bind}$ should be measured from the exterior of the ligand corona to the receptor substrate. However, if the ligands are long and flexible, or at a relatively low surface density, then $h_{bind}$ is better defined as going all the way to the solid exterior of the attacker.

\begin{figure*}
	\centering
	\includegraphics[width= 1.0\textwidth]{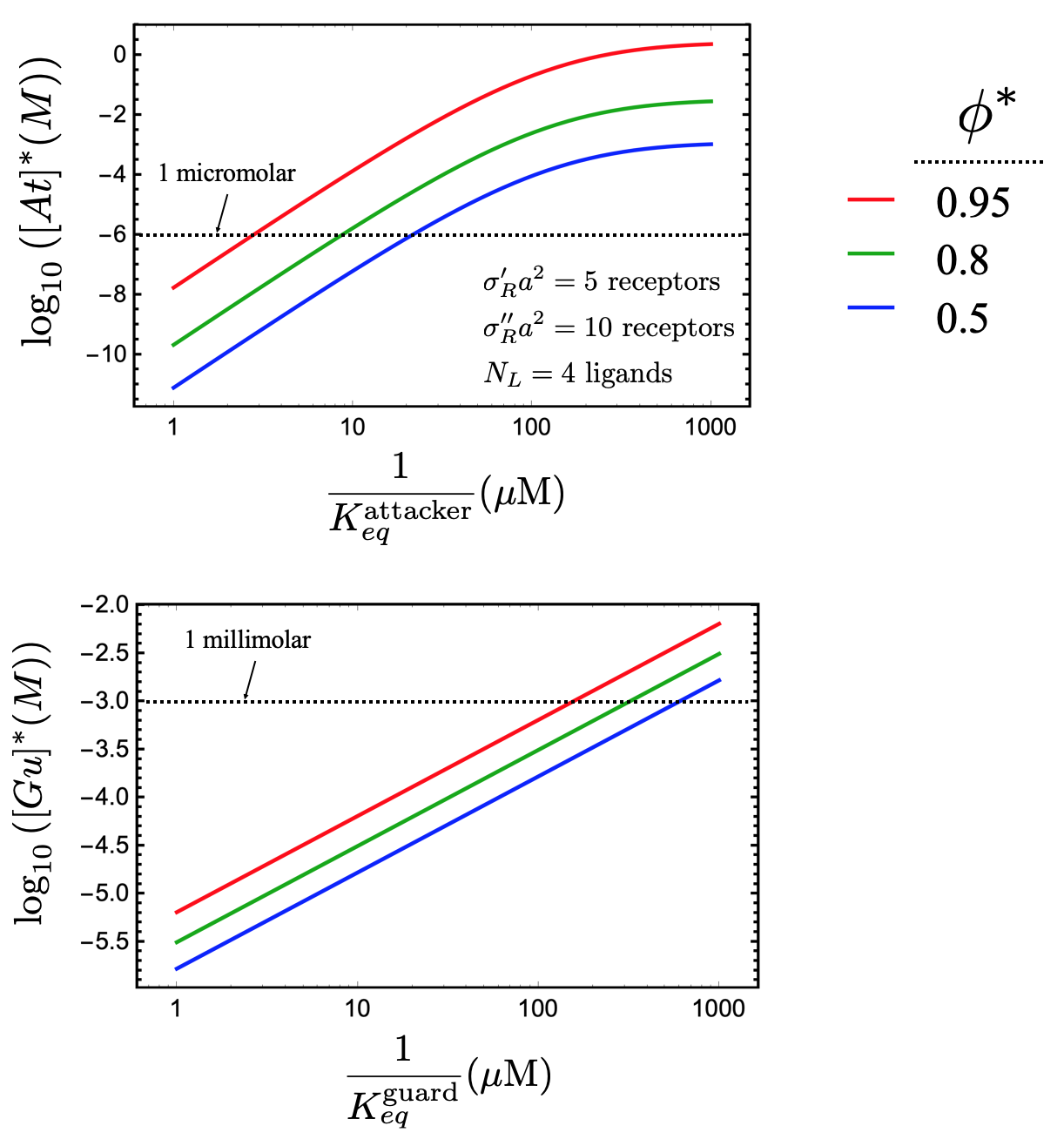}
	\caption{ Logarithm of the optimal attacker (upper panel) and guard (lower panel) concentrations as a function of reciprocal binding constants $1/K_{eq}^{\text{attacker}}$ and $1/K_{eq}^{\text{guard}}$ (in units of micromolar). Results are calculated using Eqs. \ref{eqn:AtDesign} and \ref{eqn:GuDesign}. Attackers are defined to have $N_L = 4$, a diameter of $a = 50$nm, and an equilibrium binding height $h_{bind} = 15$nm. The low- and high-receptor-density surfaces have $\sigma_R' a^2 = 5$ and $\sigma_R'' a^2 = 10$ receptors per attacker footprint $a^2$, respectively. Calculations are displayed for three choices of targeting effectiveness $\phi^* = 0.95$, $0.8$, and $0.5$ (red, green, blue). For guard calculations, attacker ligand/receptor binding constant is set to $1/K_{eq}^{\text{attacker}} = 10$ micromolar. }
	\label{fig:OptConcVsKd}
\end{figure*}  

For purpose of demonstration, Figure \ref{fig:OptConcVsKd} shows experimentally-relevant examples of how the optimal attacker and guard concentrations given by Eqs. \ref{eqn:AtDesign} and \ref{eqn:GuDesign} vary with choice of binding constants. The diameter of an attacker is set to $a = 50$nm, with an equilibrium binding height $h_{bind} = 15$nm. The low-receptor-density surface is chosen to have $\sigma_R' a^2 = 5$ receptors per attacker ``footprint'', and the high-density surface $\sigma_R'' a^2 = 10$. The attacker particles are defined to have $N_L = 4$. Results have been calculated for three choices of desired targeting effectiveness $\phi^*$.

In Eq. \ref{eqn:GuDesign}, we immediately see that the optimal guard concentration always scales with the binding constant $K_{eq}^{\text{guard}}$ as
\begin{equation}
	[Gu]^* \propto \frac{1}{K_{eq}^{\text{guard}}} \nonumber
\end{equation}
The attackers behave in a more complex way. However, when the attacker ligands are strong-binding, then the ratio $(q_L(c_R'') / q_L(c_R')) \rightarrow (c_R'' / c_R')$ in Eqs. \ref{eqn:AtDesign} and \ref{eqn:GuDesign}. In this limit, the optimal attacker concentration scales as
\begin{equation}
	[At]^* \propto \left(\frac{1}{K_{eq}^{\text{attacker}}}\right)^{N_L} \nonumber
\end{equation}
This scaling relation is noted in the upper panel of Figure \ref{fig:OptConcVsKd} when the attacker ligands are strong-binding. The logarithm of $[At]^*$ varies nearly linearly with the logarithm of the reciprocal ligand/receptor binding constant (i.e. the \emph{dissociation} constant $K_{d}^{\text{attacker}}$) when $K_{d}^{\text{attacker}}$ is small. For relatively strong-binding ligands, where $K_{d}^{\text{attacker}}$ is around 1 micromolar, then the optimal solution concentration of attackers is in the micromolar to nanomolar range. Weakening the ligand/receptor $K_{d}^{\text{attacker}}$ brings $[At]^*$ to higher concentrations. Choosing a larger targeting effectiveness $\phi^*$ also acts to increase the optimal $[At]^*$ range.

The story for the guards is similar in the lower panel of Figure \ref{fig:OptConcVsKd}. The scaling of $[Gu]^* \propto \frac{1}{K_{eq}^{\text{guard}}}$ is clear, though it is less steep than for the attackers, due to the lack of the exponent $N_L$ on the (monovalent) guards. When the guard dissociation constant is around one micromolar, the optimal concentration is also in the micromolar range. Increasing the dissociation constant into the millimolar range accordingly brings the optimal guard concentration into that range as well.

\subsection{Kinetics on the road to equilibrium $\&$ suggested sequence of ingredients}

Strong-binding attackers and guards have the limitation of long equilibration times, since both species will have strong affinity for both receptor surfaces. The desired equilibrium binding distribution of attackers and guards shown in Figure \ref{fig:AttackerGuardBinding}b may therefore take a very long time to achieve, as the unbinding rates of the attackers and guards on either surface decreases exponentially as the overall binding free energy grows larger and more negative/favourable.

For example, monovalent binders with a binding free energy of $f$ (corresponding to an equilibrium association constant of $K_{eq}$) have an unbinding timescale that goes as the Arrhenius form
\begin{equation}
	\frac{\tau^{\text{mono}}_{\text{off}}}{\tau^0} = e^{-\beta f} \propto K_{eq},
\end{equation}
where $\tau^0$ is a characteristic timescale. (This form assumes that there is no appreciable activation barrier to unbinding.) The unbinding timescale $\tau^{\text{mono}}_{\text{off}}$ grows longer for strong-binding (e.g. larger negative $\beta f$, larger $K_{eq}$) particles. For multivalent binders with ligand/receptor bonds of strength $f_{\text{LR}}$, the unbinding timescale increases exponentially with the average number $\bar{m}$ of bonds: \cite{Licata:2008kp}
\begin{equation}
	\frac{\tau^{\text{multi}}_{\text{off}}(\bar{m})}{\tau^0_{\text{LR}}} = \frac{e^{-\beta G_{NS}}}{\bar{m}} e^{-\beta \bar{m} f_{\text{LR}}} \propto \left(K^{\text{LR}}_{eq}\right)^{\bar{m}},
\end{equation}
where $\tau^0_{\text{LR}}$ is the timescale (reciprocal rate) of ligand/receptor association when both entities are free in solution at 1 molar reference concentration, and $\beta G_{NS}$ contains all of the non-specific interaction free energy contributions between the multivalent particle and the target surface as noted earlier in Eq. \ref{eqn:MVBindingFreeEnergyAllTerms}.

\begin{figure*}
	\centering
	\includegraphics[width= 0.90\textwidth]{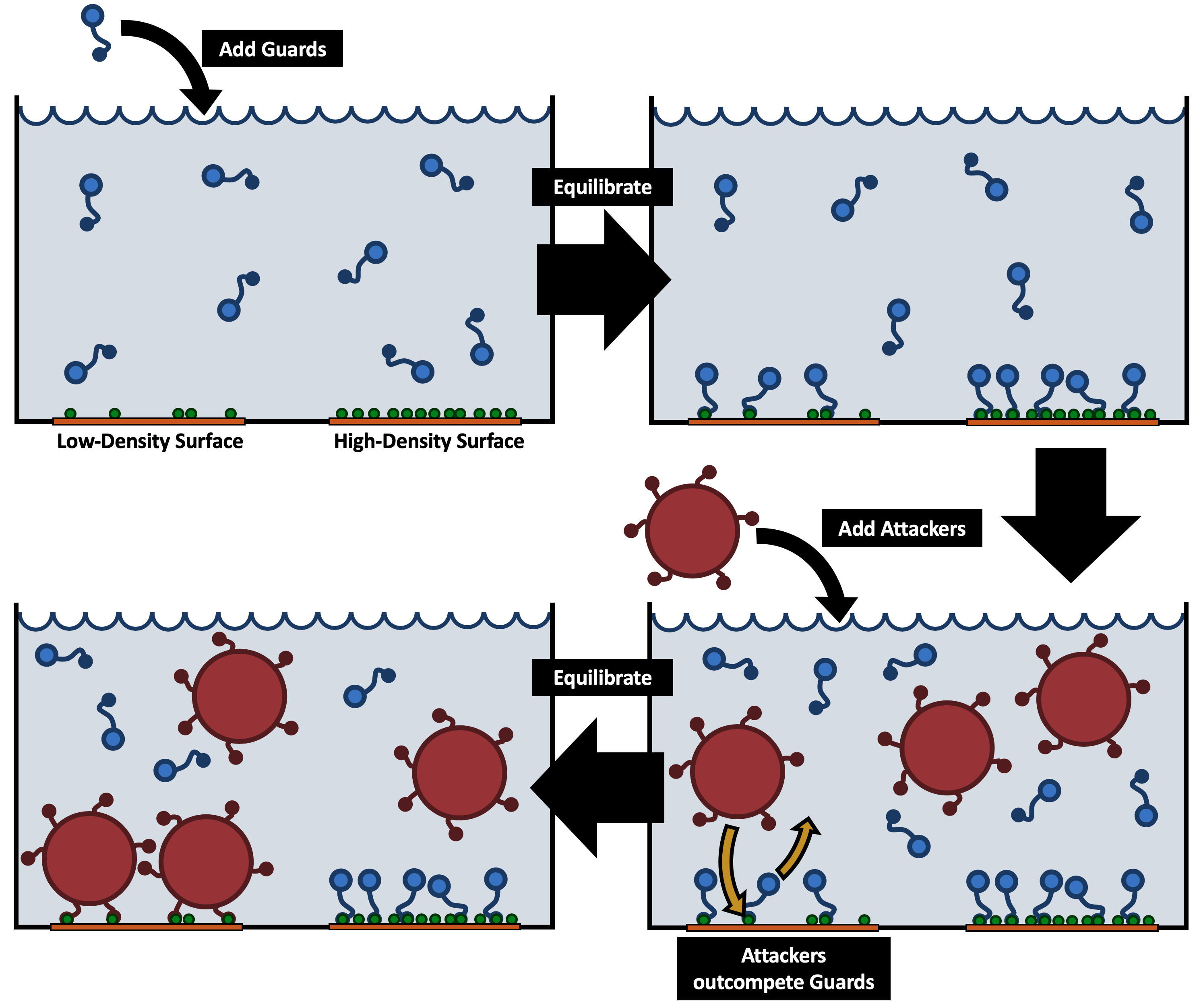}
	\caption{ Graphical depiction of a kinetically-facile approach to reaching Attacker $\&$ Guard equilibrium in a hypothetical solution containing a low- and high-receptor-density surface. Adding guards (top left) to the solution and then equilibrating leads guards to bind to receptors on both surfaces (upper right). Attackers are then titrated into the solution (lower right), and they outcompete for binding on the low-density surface due to their more favourable binding free energy at that receptor density (lower left).  }
	\label{fig:RecipeCartoon}
\end{figure*}

When the ligands on the attackers are strong-binding, as the present strategy calls for, then $\tau^{\text{multi}}_{\text{off}}(\bar{m})$ can potentially grow very long compared to $\tau^{\text{mono}}_{\text{off}}$ for the guards. To pre-emptively circumvent this kinetic barrier, we can envision the recipe depicted in Figure \ref{fig:RecipeCartoon} for reaching an attacker $\&$ guard binding equilibrium:
\begin{enumerate}
	\item{Add monovalent guards.}
	\item{Equilibrate.}
	\item{Add multivalent attackers.}
	\item{Equilibrate again.}
\end{enumerate}
By this route, the only exchange necessary is on the low-receptor-density surface, where monovalent guards must unbind in order to allow the more favourably-binding multivalent attackers to attach (Figure \ref{fig:RecipeCartoon}, lower right).  In experimental design, the guard $K_{eq}^{\text{guard}}$ can be chosen to be a strong-binding yet kinetically-reasonable value, and then the concentration $[Gu]^*$ can be chosen via Eq. \ref{eqn:GuDesign}. The guard binding constant will therefore set the timescale $\tau_{\text{off}}$ that must be waited for the final equilibrium to be reached.

\subsection{Making a more tolerant design}

In Eq. \ref{eqn:OptTolerance}, we found that the tolerance of the Attacker \& Guard strategy grows exponentially larger by designing a larger (more negative) free energy gap $\beta \Delta G^*$. Let us now examine in detail how $\beta \Delta G^*$ depends on the attacker and guard parameters, assuming that we always choose the optimal concentrations given by Eqs. \ref{eqn:AtDesign} and \ref{eqn:GuDesign}.

Equations \ref{eqn:FEGuards} and \ref{eqn:FEAttackers} can be used to write two equivalent equations for $\beta \Delta G^*$ by invoking the optimisation condition found in Eq. \ref{eqn:OptimalCondition}. These are: $(\beta \Delta G^*)_1 = \beta G_{\text{attacker}}(N_R') - \beta G_{\text{guard}}(N_R')$; and $(\beta \Delta G^*)_2 = \beta G_{\text{guard}}(N_R'') - \beta G_{\text{attacker}}(N_R'')$. Given that $[(\beta \Delta G^*)_1 + (\beta \Delta G^*)_2]/2 = \beta \Delta G^*$ by definition, then we arrive at
\begin{equation}
	\beta \Delta G^* = \frac{N_L}{2} \ln{\left(\frac{q_L(c_R'')}{q_L(c_R')}\right)} - C_{\text{guard}}^* \left(\frac{N_A a^2(c_R'' - c_R')}{2}\right).
	\label{eqn:FreeEnergyGapVariation}
\end{equation}
The factor of Avogadro's number $N_A$ is necessary in the second term, in order to properly convert the surface receptor molarities $c_R'$ and $c_R''$ into particle counts.  In terms of experimental units, the guard design parameter $C_{\text{guard}} = \ln{\left(1 + [Gu] K_{eq}^{\text{guard}}\right)}$; this is numerically identical to the statistical-mechanical definition in Eq. \ref{eqn:GuardParameter}. For further discussion on this equivalence, refer back to Eq. \ref{eqn:KeqFromStatMech} in Appendix \ref{app:OptimalDesign}. 

Equation \ref{eqn:FreeEnergyGapVariation} has two distinct terms. The first is a positive (unfavourable) contribution that depends on the attacker design parameters $K_{eq}^{\text{attacker}}$ (in the $q_L(c_R)$ factors) and $N_L$ (in the prefactor). The second term is a negative (favourable) contribution that depends on the guard design $C_{\text{guard}}$.

These two terms can be analysed in the context of Figure \ref{fig:AttackerGuardBinding}a. Clearly, the only way to \emph{increase} the size of the free energy gap $\beta \Delta G^*$ is to make the slope of the blue (guard) free energy curve more negative. This corresponds to choosing a larger $C_{\text{guard}}$ (either by choosing a larger guard concentration, or larger binding constant $K_{eq}^{\text{guard}}$). Doing so increases the potential effectiveness $\phi$ of the recipe in Eq. \ref{eqn:OptSelectivity}, and also the tolerance to attacker concentration variations around the optimum value.

The attackers, on the other hand, have a less obvious influence on the size of the gap. For extremely weak-binding ligands, then the ratio $q_L(c_R'') / q_L(c_R')$ in Eq. \ref{eqn:FreeEnergyGapVariation} approaches unity, causing that term to vanish to zero so that $\beta \Delta G^*$ is more negative. However, this limit is not experimentally realistic. On the other hand, making the ligands stronger-binding saturates the ratio $(q_L(c_R'') / q_L(c_R')) \rightarrow (c_R'' / c_R')$ as noted previously. It is then the number of ligands $N_L$ on the attacker that serves to multiply the logarithm of this ratio in Eq. \ref{eqn:FreeEnergyGapVariation}, suggesting that attackers with more ligands lead to a potentially less effective and less tolerant design. We return to this point shortly in more quantitative terms.

A beneficial side-effect of Eq. \ref{eqn:FreeEnergyGapVariation} is that the targeting effectiveness $\phi$ is not particularly sensitive to attacker concentration variations around the optimal value given by Eq. \ref{eqn:AtDesign}. This can be seen in the numerical examples in Figure \ref{fig:OptConcVsKd} (upper panel). The attacker concentration only shifts the binding free energy of the attackers by a constant logarithmic factor $\ln{[At]}$ (appearing as $\ln{z_{\text{attacker}}}$ in Eq. \ref{eqn:FEAttackers}). For example, shifting the attacker binding free energy (red) curve in Figure \ref{fig:AttackerGuardBinding}a downward by, say, $2 kT$, corresponds to increasing the attacker particle concentration by a large factor of $e^2 \approx 7.4$. However, this will have little impact on the effectiveness $\phi$, since increasing $\beta \Delta G(N_R')$ by $-2 kT$ and $\beta \Delta G(N_R')$ by $2 kT$ leads both to still be very near $\beta \Delta G^*$ (assuming that $\beta \Delta G^*$ is already somewhat large and negative).  Results in the upper panel of Figure \ref{fig:OptConcVsKd} illustrate this point nicely. Varying the attacker molar concentration by a factor of $10^2$ (i.e. from the red to the green dataset in that figure) only corresponds to a change in effectiveness from $\phi^* = 0.95$ to $0.8$.

In contrast, the targeting effectiveness is more sensitive to variations in guard concentration $[Gu]$, as this factor goes into the \emph{slope} of the guard binding free energy in Figure \ref{fig:AttackerGuardBinding}a (blue curve) as $z_{\text{guard}}$ in Eq. \ref{eqn:FEGuards}. For example, in the lower panel of Figure \ref{fig:OptConcVsKd}, changing the guard concentration only by a factor of two (i.e. going from the red curve to the green curve) corresponds to the same change in effectiveness from $\phi^* = 0.95$ to $0.8$ enacted by changing the attacker concentration by a factor of $100$.

\begin{figure*}
	\centering
	\includegraphics[width= 1.0\textwidth]{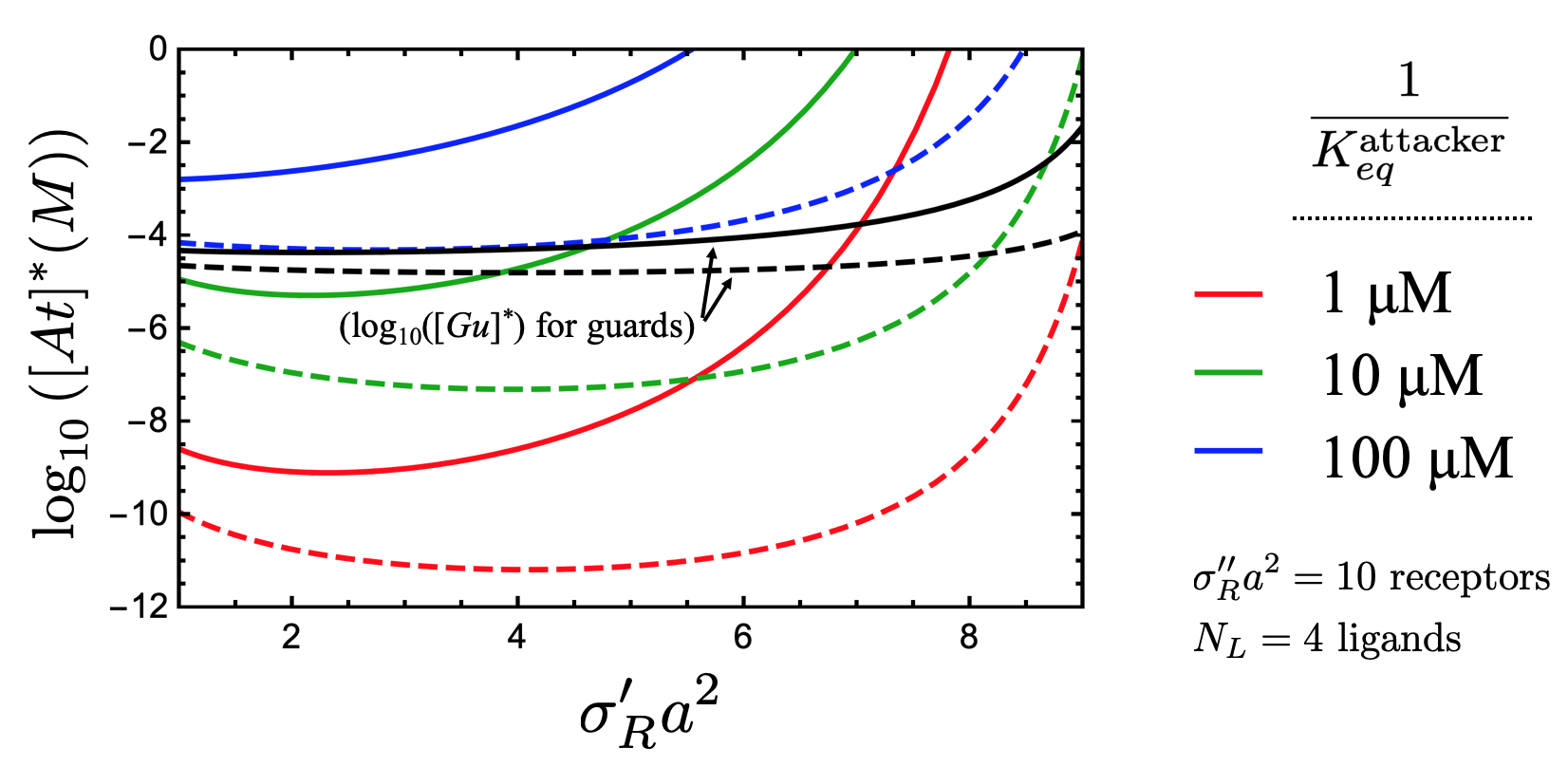}
	\caption{ Logarithm of the optimal attacker concentration $[At]^*$ as a function of the number of receptors per attacker footprint, $\sigma_R' a^2$, on the low-density surface. The high-density surface is fixed at $\sigma_R'' a^2 = 10$. Attackers have $N_L = 4$ ligands, diameter $a = 50$nm, and an equilibrium binding height $h_{bind} = 15$nm. Results for three choices of ligand/receptor dissociation constants $1/K_{eq}^{\text{attacker}}$ in the micromolar range are given (blue, green, and red). Black curves are the logarithm of the optimal guard concentration $[Gu]^*$ vs. $\sigma_R' a^2$ when their dissociation constant is $1/K_{eq}^{\text{guard}} = 10$ micromolar and when attackers also have $1/K_{eq}^{\text{attacker}} = 10$ micromolar. For all datasets, solid lines are for targeting effectiveness $\phi^* =  0.95$, and dashed lines are for $0.5$. }
	\label{fig:OptAtConcSig}
\end{figure*} 

\begin{figure*}
	\centering
	\includegraphics[width= 1.0\textwidth]{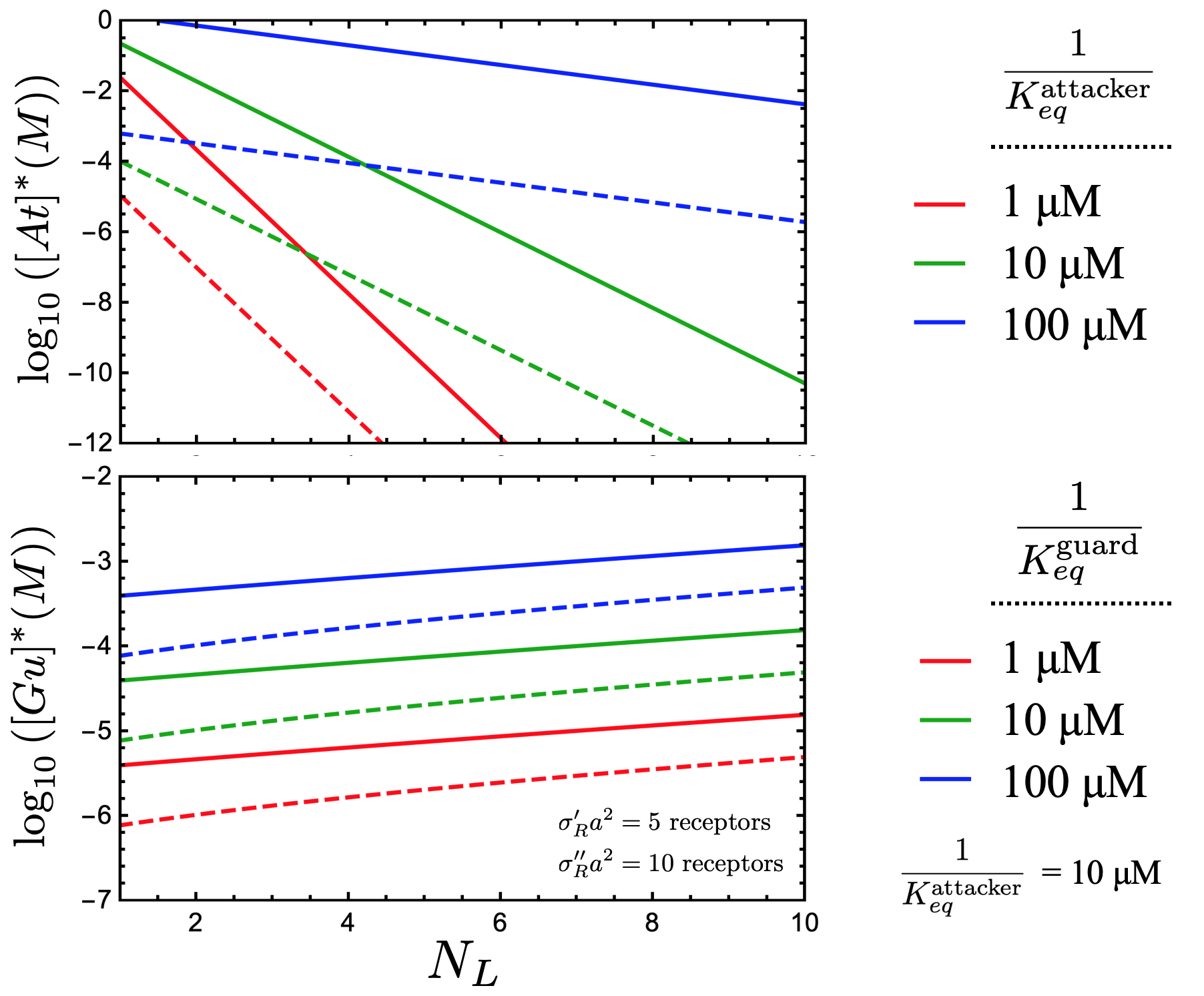}
	\caption{ Logarithm of the optimal attacker (top panel) and guard (bottom panel) concentrations as a function of the number of ligands $N_L$ on the attackers. Attackers have diameter $a = 50$nm and an equilibrium binding height $h_{bind} = 15$nm. The low- and high-receptor-density surfaces have $\sigma_R' a^2 = 5$ and $\sigma_R'' a^2 = 10$ receptors per attacker footprint $a^2$, respectively. Results are presented for three choices of micromolar-range dissociation constants $1/K_{eq}^{\text{attacker}}$ and $1/K_{eq}^{\text{guard}}$ (red, green, blue), and for two choices of targeting effectiveness $\phi^*$: 0.95 (solid) and 0.5 (dashed). For guard calculations in lower panel, $1/K_{eq}^{\text{attacker}} = 10$ micromolar. }
	\label{fig:OptAtConcNL}
\end{figure*} 

The number of receptors $N_R'$ and $N_R''$ per lattice site on the two surfaces also plays a role in $\beta \Delta G^*$. In particular, as the disparity between $N_R'$ and $N_R''$ grows larger, an attacker design with weaker binding or lower concentration can be employed in order to obtain a given targeting effectiveness $\phi^*$.  Numerical examples of this are given in Figure \ref{fig:OptAtConcSig}, showing how $[At]^*$ varies as the receptor concentration $\sigma_R'$ on the low-density surface grows closer to that on the high-density surface (having $\sigma_R'' a^2 = 10$).

A qualitatively similar trend is observed for the selections of ligand/receptor binding constant $K_{eq}^{\text{attacker}}$ and targeting effectiveness $\phi^* = 0.95$ and $0.5$. As $\sigma_R'$ grows closer to $\sigma_R''$, the attacker and guard recipe demands a larger attacker binding constant or bulk concentration. When lower targeting effectiveness is sought, then lower values for these two parameters are called for. Figure \ref{fig:OptAtConcSig} also shows how the optimal guard concentration $[Gu]^*$ varies comparatively little with $\sigma_R'$ at fixed $\sigma_R''$, assuming a moderate-binding $1/K_{eq}^{\text{guard}} = 10$ micromolar.

A somewhat counter-intuitive observation in Eq. \ref{eqn:FreeEnergyGapVariation} is that increasing the number of ligands $N_L$ on the attackers actually leads to a \emph{smaller} free energy gap. Looking at Figure \ref{fig:AttackerGuardBinding}a, a larger $\beta \Delta G^*$ is obtained when the attacker free energy (red curve) approaches behaving like a horizontal line between $N_R'$ and $N_R''$. However, increasing $N_L$ causes the attacker free energy to exhibit a more negative gradient for larger values of $N_R$, as demonstrated in Figure \ref{fig:MVBindingVsNR}b. Larger attacker valence therefore limits the targeting effectiveness of the attacker/guard recipe, again in contrast to standard multivalent reasoning. On the other hand, choosing small $N_L$ has kinetic consequences, as attackers with fewer ligands will exhibit a slower rate of surface adsorption.

Figure \ref{fig:OptAtConcNL} presents numerical results for how the optimal concentrations $[At]^*$ and $[Gu]^*$ vary with $N_L$. Examining Eq. \ref{eqn:AtDesign} and taking the strong-ligand limit where $(q_L(c_R'') / q_L(c_R')) \rightarrow (c_R'' / c_R')$ yields a scaling of
\begin{align}
	\ln{[At]^*} \propto &-N_L \ln{\left(\frac{K_{eq}^{\text{attacker}} \sqrt{c_R' c_R''}}{h_{bind}}\right)} \nonumber \\
	&+ \frac{N_L}{2} \left(\frac{c_R'' + c_R'}{c_R'' - c_R'}\right) \ln{\left(\frac{c_R''}{c_R'}\right)}.
\end{align}
The first term in this expression tends to dominate in Figure \ref{fig:OptAtConcNL} (upper panel), giving rise to the linear trends. As intuition would suggest, adding more ligands with the same binding strength $K_{eq}^{\text{attacker}}$ to the attackers leads to a lower optimal solution concentration $[At]^*$. Increasing the intended targeting effectiveness $\phi^*$ just causes a uniform upward shift in the optimal concentration $[At]^*$. However, the optimal guard concentration $[Gu]^*$ changes little with $N_L$ in Figure \ref{fig:OptAtConcNL} (lower panel). In the strong-ligand limit for the \emph{attackers}, the optimal \emph{guard} concentration in Eq. \ref{eqn:GuDesign} scales as
\begin{equation}
	\ln{[Gu]^*} \propto \frac{N_L}{a^2 \left(c_R'' - c_R'\right)} \ln{\left(\frac{c_R''}{c_R'}\right)}.
\end{equation}
This lacks the strong binding constant dependence like the attackers.

\subsection{Complications to effective targeting, and to the present theory}

It has been assumed that the attackers are large enough when bound so as to sterically exclude all receptors in their surface ``footprint'' of area $a^2$ from binding to any guard particles. This might be difficult to achieve in experiment; guard particles might slip beneath the bound attackers to occupy some of the receptors, thereby reducing the statistical average number of attackers bound on the target low-density surface. The quantitative result in Figure \ref{fig:AttackerGuardBinding}b would be a lowering of the red curve and a raising of the blue curve in the low-receptor range (i.e. on the surface(s) where attackers dominate). The overall targeting effectiveness $\phi$ (by Eq. \ref{eqn:Effectiveness}) will be accordingly reduced. In general, the more easily that guards can penetrate beneath bound attackers, the lower the resulting effectiveness of the strategy will be. One possibility is to add bulky groups to the guards, so that they are less likely to invade attacker-occupied surface territory.

In addition to ligand-receptor interactions, it was noted in Eq. \ref{eqn:MVBindingFreeEnergyAllTerms} that  multivalent particles can also have a non-specific interaction free energy $\beta G_{NS}$ with their target surface. However, as demonstrated in Eq. \ref{eqn:MVBindingFE}, any such contribution just becomes an additive factor in the multivalent binding free energy. Therefore, the only influence of $\beta G_{NS}$ is to shift the attacker binding free energy (e.g. red curve, Figure \ref{fig:AttackerGuardBinding}a) by a constant factor. This will quantitatively shift the predicted $[At]^*$ upward if $\beta G_{NS}$ is positive, or downward if $\beta G_{NS}$ is negative. However, it will not affect the possibility of selectively targeting the low-density surface.

Early in our discussion, we made the approximation that the attacker binding free energy follows the form of Eq. \ref{eqn:MVBindingFE}, so that we could proceed analytically. This expression \emph{over-estimates} the binding free energy for a multivalent particle / surface interaction in which the number of receptors within binding range of the particle is near or less than the number of ligands on the particle, particularly if the ligands are strong-binding. However, the molecular recipe arising out of our discussion is more selective when the multivalent attackers have \emph{few} ligands (as opposed to many). In cases where the number of receptors within binding range of the attacker is small relative to the number of ligands on the particle, then the true binding free energy will be less favourable (less negative) than predicted by Eq. \ref{eqn:MVBindingFE}. The predicted optimal attacker concentration $[At]^*$ will therefore be underestimated by Eq. \ref{eqn:AtDesign} for the system at hand. This quantitative difference does not prevent selective targeting of the low-receptor-density surface, nor does it qualitatively alter the targeting recipe presented here.

\section{Conclusions \& Experimental Implementation}

Multivalent particles cannot, on their own, selectively bind to a surface with a low density of receptors, while not binding to one with a higher density of the same receptors in the same system. To address this challenge, we have defined a strategy using competitive binding of \emph{two} particle species. The first species, called ``guards'', are monovalent particles that bind equally well to any receptor on any surface. A second species, called ``attackers'', are multivalent particles designed to outcompete the guards for binding on the low-receptor-density surface, but not on the surface with higher receptor density. At equilibrium, therefore, the attackers occupy the low-density surface, while the guards occupy the high-density surface.

An optimal targeting recipe, and several guidelines, have been derived and deduced in our discussion that can be directly employed in experiment. These are now summarised:
\begin{itemize}
	\item{The guards are small (receptor-sized) monovalent particles, with a receptor association constant of $K_{eq}^{\text{guard}}$, and a solution concentration of $[Gu]$. These particles are added \emph{first} in the system, and allowed to bind and equilibrate on both receptor surfaces.}
	\item{The attackers are larger multivalent particles at a solution concentration of $[At]$, each having $N_L$ ligands that can reach multiple receptors when the host multivalent particle is at a given fixed surface position. Their ligand-receptor association constant is $K_{eq}^{\text{attacker}}$. These particles are added \emph{second} into the system. The attackers will, by design, outcompete the guards for binding on the low-receptor-density surface, but not the high-receptor-density surface.}
	\item{Equations \ref{eqn:AtDesign} and \ref{eqn:GuDesign} define the optimal attacker and guard solution concentrations, given their chosen individual ligand-receptor binding constants $K_{eq}^{\text{attacker}}$ and $K_{eq}^{\text{guard}}$, the attacker valence $N_L$, and the surface receptor molarities $c_R'$ and $c_R''$ on the low- and high-receptor-density surfaces, respectively. These equations are employed by inputting a desired ``targeting effectiveness'' parameter $\phi^*$ between $1$ (perfect attacker binding selectivity for the low-receptor-density surface) and $0$ (no binding selectivity).}
	\item{More effective targeting ($\phi^*$ approaching $1$) requires \emph{stronger-binding} attackers and guards, at \emph{larger} solution concentrations. This is contrary to standard multivalent targeting. However, stronger-binding guards also lead to longer equilibration time on the low-density surface when the attackers are added. The larger the difference between the receptor densities on the two surfaces, the weaker the attacker and guard binding strengths/concentrations can be in order to achieve a given targeting effectiveness.}
	\item{More effective targeting occurs when the number of ligands on the attackers is small. However, multivalent particles with fewer ligands will have a longer timescale for forming bonds with surface receptors.}
\end{itemize}
The best approach is therefore to use strong-binding guards that still have a reasonable unbinding timescale, and comparably strong-binding attackers that have several ligands. The best design can be identified by exploring a range of targeting effectiveness values $\phi^*$ in Eqs. \ref{eqn:AtDesign} and \ref{eqn:GuDesign}, to find arrive at a guard/attacker motif that is kinetically and thermodynamically suitable. After putting the guards into the system at a concentration of $[Gu]^*$ based on Eq. \ref{eqn:GuDesign}, attackers can be titrated in until they are near a concentration of $[At]^*$ given by Eq. \ref{eqn:AtDesign}. Values of $[At] > [At]^*$ may work equally well, in order to overcome additional non-specific binding free energy contributions between the attackers and the target surface; the effectiveness of targeting does not depend strongly on $[At]$ above $[At]^*$.

A prime application for this targeting strategy is to selectively image cell surfaces, or regions of cell surfaces, with locally low receptor density compared to other cell surfaces in the same system \emph{in vitro}. This could be done by attaching a fluorescent probe to the attackers, but not the guards. Examples of promising structures that could act as attackers include ligand-coated nanoparticles, functionalised vesicles, DNA origami / dendrimer constructs, star polymers, or modified viruses. The binding equilibrium constants of the guards and attackers could be tuned by a DNA approach, e.g. like in DNA-PAINT.\cite{Delcanale:2018fy} The attacker and guard recipe could also be used to selectively sequester or aggregate a population of nanoscopic entities in solution with a low receptor density, with attackers acting as the aggregation/sequestration agent.

\section{Acknowledgments}

This work has been carried out whilst financially supported by the European Research Council, under the European Union's Seventh Framework Programme (FP/2007-2013) / ERC Grant Agreement no. 607602 (``SASSYPOL''), as well as the Netherlands 4TU.High-Tech Materials research programme `New Horizons in designer materials' (www.4tu.nl/htm). I wish to thank Tine Curk, Jurriaan Huskens, Wouter Ellenbroek, and Cornelis Storm for helpful discussions on this work, as well as Bart Markvoort, Bortolo Mognetti, Stefano Angioletti-Uberti, and Daan Frenkel for critical readings of the manuscript.

\appendix

\section{Multivalent attacker and monovalent guard binding probabilities}
\label{app:MonovalentGuardBinding}

The partition function for a surface lattice site is given, based on Eqs. \ref{eqn:FEGuards} and \ref{eqn:FEAttackers}, as
\begin{align}
	Q(N_R) &= \left(1 + z_{\text{guard}} e^{-\beta f_{\text{guard}}}\right)^{N_R} \nonumber \\
	& + z_{\text{attacker}} \left(1 + N_R e^{-\beta f_{\text{attacker}}}\right)^{N_L} \nonumber \\
	&= Q_{\text{guard}}(N_R) + Q_{\text{attacker}}(N_R).
\end{align}
The first term in $Q(N_R)$ represents all possible guard binding states, and the second term is for all attacker binding states. The state in which the lattice site has neither an attacker nor any guards bound, having a weight of unity, is included in the first term of $Q(N_R)$. This partition function also includes the state in which an attacker is within the surface lattice site, but has no ligands bound to receptors.

The probability that a surface lattice site is occupied by an attacker is just the ratio of $Q_{\text{attacker}}(N_R)$ to $Q(N_R)$ for a given number of receptors $N_R$ in the lattice site:
\begin{align}
	P_{b}^{\text{attacker}}(N_R) &= \frac{Q_{\text{attacker}}(N_R)}{Q_{\text{attacker}}(N_R) + Q_{\text{guard}}(N_R)} \nonumber \\
	&= \frac{e^{-\beta G_{\text{attacker}}(N_R)}}{e^{-\beta G_{\text{attacker}}(N_R)} + e^{-\beta G_{\text{guard}}(N_R)}}.
\end{align}
By defining the quantity
\begin{equation}
	\beta \Delta G(N_R) = \beta G_{\text{attacker}}(N_R) - \beta G_{\text{guard}}(N_R),
\end{equation}
then $P_{b}^{\text{attacker}}(N_R)$ reduces to the simple form
\begin{equation}
	P_{b}^{\text{attacker}}(N_R) = \frac{e^{-\beta \Delta G(N_R)}}{1 + e^{-\beta \Delta G(N_R)}}
\end{equation}
as shown in Eq. \ref{eqn:AttackerBindingProb}.

The probability that a single receptor on the surface is occupied by a monovalent guard is found by
\begin{equation}
	P_{b}^{\text{guard}}(N_R) = \frac{1}{N_R} \frac{d \ln{Q(N_R)}}{d \beta \mu_{\text{guard}}},
\end{equation}
where $\beta \mu_{\text{guard}} = \ln{z_{\text{guard}}}$ is the chemical potential of the guards in solution. This leads to
\begin{equation}
	P_{b}^{\text{guard}}(N_R) = \left( \frac{z_{\text{guard}} e^{-\beta f_{\text{guard}}}}{1 + z_{\text{guard}} e^{-\beta f_{\text{guard}}}}\right) \left(1 - P_b^{\text{attacker}} (N_R)\right),
\end{equation}
shown as Eq. \ref{eqn:GuardBindingProb} in the main text.

\section{Optimal targeting effectiveness and tolerance}
\label{app:SelectivityAndTolerance}

The targeting effectiveness is defined as
\begin{equation}
	\phi \equiv P_b^{\text{attacker}}(N_R') - P_b^{\text{attacker}}(N_R''),
	\label{eqn:AppSelectivity}
\end{equation}
where
\begin{equation}
	P_b^{\text{attacker}} (N_R) = \frac{e^{-\beta \Delta G(N_R)}}{1 + e^{-\beta \Delta G(N_R)}}.
\end{equation}
The free energy difference $\beta \Delta G(N_R)$ is

\begin{align}
	&\beta \Delta G(N_R) = \beta G_{\text{attacker}}(N_R) - \beta G_{\text{guard}}(N_R) \nonumber \\
	&=  -N_L \ln{\left[q_L(N_R)\right]} - \ln{z_{\text{attacker}}} + N_R C_{\text{guard}},
	\label{eqn:TotalFreeEnergyAppendix}
\end{align}	
where
\begin{equation}
	q_L(N_R) = \left(1 + N_R e^{-\beta f_{\text{attacker}}}\right)
	\label{eqn:LigPartitionFunction}
\end{equation}
is the partition function for one ligand on the attacker given $N_R$ possible receptors to attach to nearby.

Choosing the parameter $C_{\text{guard}}$ effectively sets the guard binding free energy and fugacity in solution. In experiment, it is also reasonable to assert that the number of ligands $N_L$ and ligand/receptor binding free energy $f_{\text{attacker}}$ have been set based on the chemical design of the attackers. Thus, the remaining free variable to optimize is the attacker fugacity $z_{\text{attacker}}$---that is, the solution concentration $[At]$ of attackers that maximises the targeting effectiveness $\phi$.

The attacker fugacity yielding the largest possible effectiveness $\phi$ is obtained where
\begin{align}
	&\frac{d \phi}{d \ln{z_{\text{attacker}}}} = \frac{d (\beta \Delta G(N_R''))}{d \ln{z_{\text{attacker}}}} \frac{e^{-\beta \Delta G(N_R'')}}{\left(1 + e^{-\beta \Delta G(N_R'')}\right)^2} \nonumber \\
	& - \frac{d (\beta \Delta G(N_R'))}{d \ln{z_{\text{attacker}}}}  \frac{e^{-\beta \Delta G(N_R')}}{\left(1 + e^{-\beta \Delta G(N_R')}\right)^2} = 0.
	\label{eqn:FindingDerivativeAppendix}
\end{align}
Inspecting Eq. \ref{eqn:TotalFreeEnergyAppendix}, the two derivatives of $\beta \Delta G(N_R)$ with respect to $\ln{z_{\text{attacker}}}$ are identical and independent of $N_R$. Thus, Eq. \ref{eqn:FindingDerivativeAppendix} is zero when either $\beta \Delta G(N_R') = \beta \Delta G(N_R'')$, or $\beta \Delta G(N_R') = -\beta \Delta G(N_R'')$. The first solution yields an effectiveness of zero via Eq. \ref{eqn:AppSelectivity}, which is obviously not the solution we want. The second solution,
\begin{equation}
	\beta \Delta G(N_R') = -\beta \Delta G(N_R'') \equiv \beta \Delta G^*,
\end{equation}
inserted into Eq. \ref{eqn:AppSelectivity} yields the optimal effectiveness
\begin{equation}
	\phi^* \equiv \frac{e^{-\beta \Delta G^*} - 1}{e^{-\beta \Delta G^*} + 1}.
\end{equation}
This is given as Eq. \ref{eqn:OptSelectivity} in the main text.

To ensure that $\beta \Delta G(N_R') = -\beta \Delta G(N_R'')$ corresponds to a maximum in Eq. \ref{eqn:AppSelectivity}, we check the sign of the second derivative of  the function $\phi$ with respect to $\ln{z_{\text{attacker}}}$ at $\beta \Delta G(N_R') = -\beta \Delta G(N_R'') = \beta \Delta G^*$:
\begin{align}
	\frac{d^2 \phi}{\left[d \ln{z_{\text{attacker}}}\right]^2} &= 2 e^{-\beta \Delta G^*} \left[\frac{1 - e^{-\beta \Delta G^*}}{\left(1 + e^{-\beta \Delta G^*}\right)^3}\right].
\end{align}
The second derivative is always negative, and therefore the condition $\beta \Delta G(N_R') = -\beta \Delta G(N_R'')$ always corresponds to a maximum, as long as $\beta \Delta G^* < 0$. This is true for \emph{any} design which selectively targets the attackers to the low-receptor-density surface. When $\beta \Delta G^* << 0$, then
\begin{align}
	-\left(\frac{d^2 \phi}{\left[d \ln{z_{\text{attacker}}}\right]^2}\right)^{-1} &\approx \frac{1}{2} e^{-\beta \Delta G^*} \nonumber \\
	&\equiv \text{Design Tolerance}.
\end{align}

This illustrates how designing the attackers and guards to have a more negative $\beta \Delta G^*$ causes the design to be more robust/tolerant to variations in molecular construction and concentration. 

\section{Optimal attacker and guard design parameters}
\label{app:OptimalDesign}

The binding free energies for guards ($\beta G_{\text{guard}}(N_R)$) and attackers ($\beta G_{\text{attacker}}(N_R)$) are given by Eqs. \ref{eqn:FEGuards} and \ref{eqn:FEAttackers}, respectively. Guards are defined by the choice of the parameter $C_{\text{guard}}$, and attackers are given a pre-defined valence $N_L$ and ligand/receptor binding free energy $f_{\text{attacker}}$. In Appendix \ref{app:SelectivityAndTolerance}, we showed that the choice of attacker fugacity $z_{\text{attacker}}$ that maximises the effectiveness $\phi$ corresponds to when $\beta G_{\text{attacker}}(N_R') - \beta G_{\text{guard}}(N_R') = \beta G_{\text{guard}}(N_R'') - \beta G_{\text{attacker}}(N_R'')$.  With Eqs. \ref{eqn:FEGuards} and \ref{eqn:FEAttackers}, this condition yields the  equation  
\begin{align}
	&\ln{z_{\text{attacker}}} = \frac{(N_R' + N_R'')}{2} C_{\text{guard}}-\frac{N_L}{2} \ln{\left[q_L(N_R') q_L(N_R'')\right]}
	\label{eqn:OptimalDesignMidway}
\end{align}
where $q_L(N_R)$ is given by Eq. \ref{eqn:LigPartitionFunction}.

Next, the relation $\beta \Delta G^* = -\ln{\left(\frac{1 + \phi^*}{1 - \phi^*}\right)} = \beta G_{\text{attacker}}(N_R') - \beta G_{\text{guard}}(N_R')$ from Eq. \ref{eqn:OptSelectivity} allows us to write another equation with Eqs. \ref{eqn:FEGuards} and \ref{eqn:FEAttackers}:
\begin{equation}
	-\ln{\left(\frac{1 + \phi^*}{1 - \phi^*}\right)} = -N_L \ln{\left[q_L(N_R')\right]} - \ln{z_{\text{attacker}}} + N_R' C_{\text{guard}} 
\end{equation}
Isolating $\ln{z_{\text{attacker}}}$ in this equation and then putting it back into Eq. \ref{eqn:OptimalDesignMidway} allows us to solve for the optimal $C_{\text{guard}}$ given a choice of targeting effectiveness $\phi^*$:

\begin{align}
	&C_{\text{guard}}^* = \left(\frac{2}{N_R'' - N_R'} \right) \ln{\left[\left(\frac{q_L(N_R'')}{q_L(N_R')}\right)^{\frac{N_L}{2}} \left(\frac{1 + \phi^*}{1 - \phi^*}\right)\right]}
	\label{eqn:OptimalGuardApp}
\end{align}
Putting this back into Eq. \ref{eqn:OptimalDesignMidway} then yields
\begin{align}
	\ln{z^*_{\text{attacker}}} =&\left(\frac{N_R'' + N_R'}{N_R'' - N_R'}\right) \ln{\left[\left(\frac{q_L(N_R'')}{q_L(N_R')}\right)^{\frac{N_L}{2}} \left(\frac{1 + \phi^*}{1 - \phi^*}\right)\right]} \nonumber \\
	&-\frac{N_L}{2} \ln{\left(q_L(N_R') q_L(N_R'')\right)}.
	\label{eqn:OptimalAttackerApp}
\end{align}

Equations \ref{eqn:OptimalGuardApp} and \ref{eqn:OptimalAttackerApp} can be readily converted into a chemical equilibrium notation.    First, all instances of $N_R'$ and $N_R''$ can be identically expressed in terms of average surface receptor densities $\sigma_R'$ and $\sigma_R''$. (Recall that the receptor density $\sigma_R$ is related to $N_R$ by $N_R = a^2 \sigma_R$; the length $a$ is the diameter of one attacker particle, and $a^2$ therefore measures the area over which the particle can bind to receptors on the surface.) This substitution leads to
\begin{align}
	&C_{\text{guard}}^* = \left[\frac{2}{a^2 \left(\sigma_R'' - \sigma_R'\right)} \right] \ln{\left[\left(\frac{q_L(\sigma_R'')}{q_L(\sigma_R')}\right)^{\frac{N_L}{2}} \left(\frac{1 + \phi^*}{1 - \phi^*}\right)\right]} \label{eqn:GuardChemicalConversionStep0} \\
	&\ln{z^*_{\text{attacker}}} = \left(\frac{\sigma_R'' + \sigma_R'}{\sigma_R'' - \sigma_R'}\right)  \ln{\left[\left(\frac{q_L(\sigma_R'')}{q_L(\sigma_R')}\right)^{\frac{N_L}{2}} \left(\frac{1 + \phi^*}{1 - \phi^*}\right)\right]} \nonumber \\
	&-\frac{N_L}{2} \ln{\left(q_L(\sigma_R') q_L(\sigma_R'')\right)}.
	\label{eqn:AttackerChemicalConversionStep0}
\end{align}
where
\begin{equation}
	q_L(\sigma_R) = \left(1 + a^2 \sigma_R e^{-\beta f_{\text{attacker}}}\right)
\end{equation}
The remaining challenge is to express the following quantities in terms of the receptor binding constant $K_{eq}^{\text{attacker}}$ for ligands on the attacker, the guard binding constant $K_{eq}^{\text{guard}}$, and guard/attacker molar concentrations $[Gu]$ and $[At]$:
\begin{align}
	&C_{\text{guard}}^* =  \ln{\left(1 + z_{\text{guard}} e^{-\beta f_{\text{guard}}}\right)}^* \label{eqn:AppGuardParameter} \\
	&\ln{z^*_{\text{attacker}}} \\
	&q_L(\sigma_R) = \left(1 + a^2 \sigma_R e^{-\beta f_{\text{attacker}}}\right) \label{eqn:AppAttackerParameter}
\end{align}
This process is now described in detail.

Recall that the equilibrium constant $K_{eq}$ for two binding entities ``B'' and ``R'', in a hypothetical solution, is defined as
\begin{equation}
	K_{eq} = \frac{[BR]}{[B][R]}.
\end{equation}
Here, $[BR]$ is the equilibrium concentration of B bound to R, while $[B]$ and $[R]$ are the equilibrium concentrations of unbound B and R. Let's suppose that species $R$ are receptors on a surface, while species $B$ are binders in solution above the surface. If we go to the grand-canonical limit where the number of B particles is far in excess of the number of R particles, then $[B] \approx [B]^\circ$, where $[B]^\circ$ is the (fixed) solution concentration of B regardless of how many are bound to R. This leads to
\begin{equation}
	K_{eq} = \frac{1}{[B]^\circ} \left(\frac{[BR]}{[R]}\right) = \frac{1}{[B]^\circ} \left(\frac{N_{R,bound}}{N_{R,free}}\right)
	\label{eqn:KeqGuardsPartWay}
\end{equation}
where $N_{R,bound}$ and $N_{R,free}$ are the number of receptors that are bound to B, and unbound, respectively. The number of unbound receptors at equilibrium is given by $N_{R,free} = N_R - N_{R,bound}$, where $N_{R}$ is the total number of receptors on the surface. This is directly related to the equilibrium binding free energy $\beta f$ of B to R by the statistical mechanical relationship
\begin{equation}
	\left(\frac{N_{R,bound}}{N_R - N_{R,bound}}\right) = z([B]^\circ) e^{-\beta f},
\end{equation}
where $z([B]^\circ)$ is the fugacity corresponding to the solution concentration $[B]^\circ$. Bringing the concentration factor $[B]^\circ$ onto the left-hand side of Eq. \ref{eqn:KeqGuardsPartWay} yields
\begin{equation}
	K_{eq} [B]^\circ = z([B]^\circ) e^{-\beta f}.
	\label{eqn:KeqFromStatMech}
\end{equation}
This is the direct relationship between the experimental quantity $K_{eq} [B]^\circ$ and the statistical thermodynamic quantity $z([B]^\circ) e^{-\beta f}$.

Using Eq. \ref{eqn:KeqFromStatMech}, we can immediately convert $z_{\text{guard}} e^{-\beta f_{\text{guard}}}$ in Eq. \ref{eqn:AppGuardParameter} into experimental units for the guards:
\begin{equation}
	z_{\text{guard}} e^{-\beta f_{\text{guard}}} = [Gu] K_{eq}^{\text{guard}},
	\label{eqn:GuardChemicalConversion}
\end{equation}
where $[Gu]$ is the molar solution concentration of the guards. 

Next, the fugacity $z_{\text{attacker}}$ of the attackers is related to their molar solution concentration $[At]$ via Eq. \ref{eqn:EqForConcentrationFromFug}: 
\begin{equation}
	z_{\text{attacker}} = [At] N_A h_{bind} a^2,
\end{equation}
where $h_{bind} a^2$ is the ``localisation volume'' for placing the ligands of the particle in contact with the surface receptors.

Turning lastly to $q_L(\sigma_R)$, the quantity $\sigma_R a^2$ acts as a two-dimensional receptor fugacity and $\exp{\left(-\beta f_{\text{attacker}}\right)}$ is the binding strength term. Thus, we can again invoke Eq. \ref{eqn:KeqFromStatMech} to write Eq. \ref{eqn:AppAttackerParameter} as
\begin{equation}
	q_L([R]_{\text{eff}}) = \left(1 + [R]_{\text{eff}} e^{-\beta f_{\text{attacker}}}\right)
	\label{eqn:AttackerChemicalConversionStep4}
\end{equation}
where $[R]_{\text{eff}}$ is the effective molarity of receptors on the surface, as seen by ligands on bound attackers. The effective molarities on the low- and high-density surfaces are related to the surface (number) densities $\sigma_R'$ and $\sigma_R''$ via the equilibrium binding distance $h_{bind}$ of the attacker:
\begin{align}
	&[R]'_{\text{eff}} = \frac{\sigma_R'}{N_A h_{bind}} = \frac{c_R'}{h_{bind}} \\
	&[R]''_{\text{eff}} = \frac{\sigma_R''}{N_A h_{bind}} = \frac{c_R''}{h_{bind}}
\end{align}
For notational clarity, the molar surface receptor density $c_R = \sigma_R / N_A$ has been defined, having dimensions of moles of receptors per unit surface area.
Re-expressing Eq. \ref{eqn:AttackerChemicalConversionStep4} in terms of $c_R$ and $h_{bind}$ yields
\begin{equation}
	q_L(c_R) = \left(1 + \frac{c_R K_{eq}^{\text{attacker}}}{h_{bind}}\right)
	\label{eqn:AppLigPartFuncExpFinal}
\end{equation}

These transformations enable us to write the expressions for the optimal guard design $C^*_{\text{guard}}$ (Eq. \ref{eqn:GuardChemicalConversionStep0}) and attacker fugacity $z^*_{\text{attacker}}$ (Eq. \ref{eqn:AttackerChemicalConversionStep0}) in terms of experimental quantities:
\begin{align}
	\ln{[At]^*} = &\left(\frac{c_R'' + c_R'}{c_R'' - c_R'}\right)  \ln{\left[\left(\frac{q_L(c_R'')}{q_L(c_R')}\right)^{\frac{N_L}{2}} \left(\frac{1 + \phi^*}{1 - \phi^*}\right)\right]} \nonumber \\
	&-\frac{N_L}{2} \ln{\left(q_L(c_R') q_L(c_R'')\right)} - \ln{\left(N_A h_{bind} a^2\right)}
	\label{eqn:AttackerChemicalConversionFinal}
\end{align}
\begin{align}
	\ln{\left(1 + [Gu]^* K_{eq}^{\text{guard}}\right)}& = \left[\frac{2}{N_A a^2 \left(c_R'' - c_R'\right)} \right] \nonumber \\
	& \times \ln{\left[\left(\frac{q_L(c_R'')}{q_L(c_R')}\right)^{\frac{N_L}{2}} \left(\frac{1 + \phi^*}{1 - \phi^*}\right)\right]}.
	\label{eqn:GuardChemicalConversionFinal}
\end{align}
where $q_L(c_R)$ is calculated by Eq. \ref{eqn:AppLigPartFuncExpFinal}.

In this form, we have assumed that the ligand/receptor binding constants $K_{eq}^{\text{guard}}$ and $K_{eq}^{\text{attacker}}$, as well as the number of ligands $N_L$ on the attackers, are set based on the chemical construction of the attackers and guards. Achieving the optimum targeting effectiveness is thus left to tuning of the attacker and guard solution concentrations to the optimal values $[At]^*$ and $[Gu]^*$. The explicit equations for these two quantities are obtained by isolating $[At]^*$ and $[Gu]^*$ in Eqs. \ref{eqn:AttackerChemicalConversionFinal} and \ref{eqn:GuardChemicalConversionFinal}. This is done to yield Eqs. \ref{eqn:AtDesign} and \ref{eqn:GuDesign} in the main text.

\bibliography{main}

\end{document}